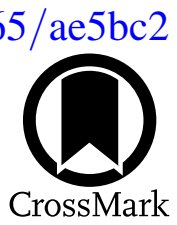

# Discovery and Characterization of White Dwarf–FGK Main-sequence Binaries within the Optical Main-sequence Locus

Mingkuan Yang (杨明宽)[1,2,3,4], Hailong Yuan (袁海龙)[1], Xiaozhen Yang (杨肖振)[1,2], Zhongrui Bai (白仲瑞)[1], Yuji He (何玉吉)[1,5], Jianping Xiong (熊建萍)[3,4], Jiao Li (李蛟)[3,4], Mengxin Wang (汪梦欣)[1], Yiqiao Dong (董义乔)[1], Ziyue Jiang (蒋子悦)[1,2], Qian Liu (刘倩)[1,2], Ganyu Li (李甘雨)[1,2], Ming Zhou (周明)[1,6], Haotong Zhang (张昊彤)[1,7], and Xuefei Chen (陈雪飞)[2,3,4,7]

[1] Key Laboratory of Optical Astronomy, National Astronomical Observatories, Chinese Academy of Sciences, Beijing 100101, People's Republic of China; htzhang@bao.ac.cn

[2] School of Astronomy and Space Science, University of Chinese Academy of Sciences, Beijing 100049, People's Republic of China; cxf@ynao.ac.cn

[3] International Centre of Supernovae (ICESUN), Yunnan Key Laboratory of Supernova Research, Yunnan Observatories, Chinese Academy of Sciences, Kunming 650216, People's Republic of China

[4] Key Laboratory for Structure and Evolution of Celestial Objects, Chinese Academy of Sciences, Kunming 650216, People's Republic of China

[5] Leibniz-Institut für Astrophysik Potsdam (AIP), An der Sternwarte 16, 14482 Potsdam, Germany

[6] College of Photonics and Optical Engineering, Aerospace Information Technology University, Jinan 250299, People's Republic of China

## Abstract

White dwarf–main sequence (WDMS) binaries provide important laboratories for studying binary evolution and the formation of low-mass white dwarfs. In this work, we identify 654 reliable WDMS candidates with FGK-type companions from an initial set of 772 ultraviolet-excess sources, selected using stellar atmospheric parameters from LAMOST spectroscopy and subsequently refined with Gaia DR3 astrometry and photometry together with ultraviolet data from GALEX. Candidates were selected based on ultraviolet excess relative to the Gaia main-sequence (MS) locus and refined using isochrone constraints to exclude systems inconsistent with MS companions. Binary spectral energy distribution fitting yields effective temperatures and radii for both components, as well as distance and extinction estimates. The MS companions are dominated by G-type stars (∼52%), with comparable fractions of F- and K-type companions, and no A-type primaries. Using white-dwarf (WD) evolutionary cooling models, we find that the WD components are predominantly low-mass ($M_{\rm WD}$ ∼ 0.2–0.4 $M_\odot$), including a substantial population of extremely low-mass (<0.3 $M_\odot$) WDs likely produced through binary interaction. The WDs are generally hot (∼$1.5 \times 10^4$ K), consistent with the ultraviolet selection bias favoring luminous, large-radius WDs. Multiepoch LAMOST radial velocities show larger amplitudes than those of a comparison sample of MS stars, supporting the close-binary nature of these systems. Although subject to strong selection effects, the catalog offers a clean and well-characterized sample of FGK +WD binaries.

## 1. Introduction

White dwarf–main sequence (WDMS) binaries are important laboratories for studying binary interaction and the formation of low-mass white dwarfs (WDs). In a primordial main-sequence (MS) binary, the initially more massive star evolves off the MS first and expands on the red giant branch (RGB) or asymptotic giant branch, eventually filling its Roche lobe and initiating mass transfer (M. de Kool 1992; B. Willems & U. Kolb 2004; J. Farihi et al. 2010; S. O. Kepler et al. 2019). The subsequent Roche-lobe overflow (RLOF) may proceed stably or become dynamically unstable and trigger a common-envelope (CE) phase, depending on the mass ratio (J. Iben & I. Livio 1993; E. F. Milone et al. 2008; A. Bobrick et al. 2017; A. C. Rubio et al. 2025). When initiated during the RGB phase, both stable RLOF and CE ejection can remove the envelope prematurely and leave behind a helium-core WD with a mass below approximately 0.4 $M_\odot$, a mass range that cannot be produced through single-star evolution within a Hubble time (A. Rebassa-Mansergas et al. 2011; J. L. A. Nandez et al. 2015). These systems therefore provide essential constraints on mass transfer, envelope ejection, and the broader physics of close-binary evolution.

Extremely low-mass (ELM) WDs, which are helium-core remnants with $M \lesssim 0.3\,M_\odot$, represent the most extreme outcomes of interaction-driven evolution. Their progenitors, if evolving as single stars, would have MS lifetimes exceeding the age of the Universe, so ELM WDs must be formed through binary mass loss (W. R. Brown et al. 2010; I. Pelisoli et al. 2018). They are produced when a ∼1–1.5 $M_\odot$ star is stripped of its envelope through RLOF or CE evolution (X. Chen et al. 2017; Z. Li et al. 2019). EL CVn binaries provide a clear example of the stable-RLOF channel: these systems contain a pre-ELM WD with an A/F-type MS companion and originate from short-period (≲a few days) solar-type binaries undergoing early, stable, nonconservative mass transfer (P. F. L. Maxted et al. 2014; X. Chen et al. 2017). Searches for EL CVn systems further show that the A/F-type companion dominates the flux from the near-ultraviolet (NUV) to the optical, while the pre-

[7] Corresponding authors.

ELM WD contributes mainly in the ultraviolet (UV; L. Wang et al. 2020; J. Xiong et al. 2025), suggesting that additional ELM WD binaries may be hidden among WDMS systems with early-type companions and detectable only through their UV excess.

Recently, J. Li et al. (2025) and X. Pérez-Couto et al. (2025) used artificial-intelligence methods to identify WDMS binaries from Gaia DR3 XP spectra and found an apparent excess of low-mass WDs. This excess may be caused by selection effects or may reflect the intrinsic WD mass distribution. However, these XP-selected samples are dominated by M dwarf companions, consistent with earlier spectroscopic surveys (A. Rebassa-Mansergas et al. 2010; J. J. Ren et al. 2014; A. Rebassa-Mansergas et al. 2012, 2016; J. J. Ren et al. 2018), in which WDMS binaries with early-type MS stars are notably underrepresented.

Previous work (A. Rebassa-Mansergas et al. 2021; A. Santos-García et al. 2025) has shown that the majority of WDMS binaries may reside within the MS region of the Gaia optical color–magnitude diagram (CMD). Early-type MS stars are more massive and intrinsically brighter than their later-type counterparts, so they can readily outshine a WD companion in the optical. This remains true even when the WD has a relatively large radius and low mass as a consequence of binary mass transfer. As a result, many WDMS systems containing low-mass or even ELM WDs may remain undetected among binaries with luminous early-type companions. In this work, we construct a sample of WDMS candidates that lie along the Gaia MS locus and host FGK-type companions, with the goal of identifying systems that may reveal distinctive WD mass distributions.

## 2. Data Selection and Methods

### 2.1. LAMOST Catalog of A, F, G, and K Stars

The Large Sky Area Multi-Object Fiber Spectroscopic Telescope (LAMOST) is a ground-based optical survey telescope designed for wide-field spectroscopic observations (X.-Q. Cui et al. 2012). With a 5° field of view and 4000 optical fibers, LAMOST can obtain low-resolution spectra for up to 4000 targets simultaneously. This capability enables highly efficient data collection and allows for repeated observations of the same sources on different nights.

As of Data Release 11 version 1.0[8] (DR11 v1.0), LAMOST has published more than 25 million stellar spectra. DR11 includes observations collected between 2022 September and 2023 June and provides 11 associated spectroscopic parameter catalogs. In this study, we make use of the Low-Resolution Spectroscopic (LRS; $R \sim 1800$) catalog of A-, F-, G-, and K-type stars from LAMOST DR11 (hereafter the AFGK catalog), which contains atmospheric parameters for 7,774,147 stars. The spectra cover a wavelength range from 3700–9000 Å.

For the AFGK catalog, we applied the following preprocessing steps to ensure the reliability and consistency of the data:

1. We enforced a minimum signal-to-noise ratio (SNR) in the $g$ filter (`snrg`) of 15. This threshold guarantees that the atmospheric parameters provided by the catalog are of sufficient quality for subsequent scientific analysis.

[8] http://www.lamost.org/dr11/

**Table 1**
Percentiles (16th, 50th, and 84th) of the Atmospheric Parameter Uncertainties for the Selected LAMOST AFGK-type Stars

| Parameter | Unit | 16% | 50% | 84% |
|---|---|---|---|---|
| $T_{\rm eff}$ | K | 28 | 43 | 82 |
| $\log g$ | dex | 0.04 | 0.06 | 0.13 |
| [Fe/H] | dex | 0.02 | 0.04 | 0.09 |

**Table 2**
Number of Sources Remaining after Each Selection Step

| Step | Selection Criterion | Number |
|---|---|---|
| | LAMOST AFGK-type stars | 7,774,147 |
| 1 | Signal-to-noise selection ($snrg \geqslant 15$) | 6,416,597 |
| 2 | Gaia DR3 crossmatch requirement | 4,473,818 |
| | Gaia DR3 quality selection | |
| 3 | Main-sequence region selection | 3,314,050 |
| 4 | Photometric and astrometric quality selection | 2,991,693 |
| 5 | Corrected excess factor ($C^{*}$) constraint | 2,435,635 |
| 6 | RUWE constraint (< 2) | 2,342,395 |
| 7 | GALEX crossmatch requirement | 1,315,020 |
| 8 | Photometric quality selection | 15,320 |
| 9 | Ultraviolet-excess selection | 772 |
| 10 | Binary SED fitting success | 759 |
| | Removal of suspected contaminants | |
| 11 | Exclusion of X-ray sources | 744 |
| 12 | Exclusion of non-main-sequence companions | 658 |
| 13 | Exclusion of chance spatial alignments | 654 |

2. We excluded entries exhibiting anomalous values in the `gaia_source_id` (mapped to the "source_id" field in the Gaia DR3 catalog). Specifically, we removed sources with `gaia_source_id` = −9999, which indicates that no corresponding Gaia observation is available. Accurate cross identification with Gaia is essential for selecting MS sources using the Gaia CMD, as outlined in Section 2.2.
3. Each row in the AFGK catalog corresponds to a unique observation ID (`obsid`), representing an individual spectroscopic observation. However, multiple `obsid` entries may correspond to the same astrophysical source if they share the same unique source ID (`uid`). For sources with multiple spectroscopic observations of the same `uid`, we grouped the observations and adopted the mean atmospheric parameters provided by the LAMOST Stellar Parameter Pipeline (Y. Wu et al. 2011). We further excluded cases in which a single `uid` is associated with multiple distinct Gaia source identifiers, since the lower angular resolution of LAMOST compared to Gaia may lead to blended spectra and consequently unreliable atmospheric parameters. The associated uncertainties were taken as the larger of (1) the maximum reported uncertainty among the individual spectra and (2) the maximum deviation of the individual measurements from the adopted mean value. The resulting uncertainty distributions are summarized in Table 1.

Following these filtering and aggregation procedures, which constitute the initial stages of the workflow later summarized in Table 2, we derived a sample comprising 4,473,818 AFGK-type stars with reliable atmospheric parameters and valid Gaia cross identifications, ensuring a well-defined parent population for the subsequent identification of WD companions.

### 2.2. Identification of Main-sequence Components

We crossmatched the AFGK-type sources obtained from the LAMOST catalog with the Gaia DR3 database using their `source_id`, and retrieved the corresponding astrometric and photometric information. Gaia DR3 provides high-precision measurements for over 1.8 billion sources, including parallaxes and broadband photometry in the $G$, $G_{\rm BP}$, and $G_{\rm RP}$ bands (Gaia Collaboration et al. 2023). These data are essential for constructing CMDs and identifying stars located on the MS.

We used publicly available dust maps to account for reddening and extinction. The Gaia parallaxes were first corrected for the global zero-point offset following L. Lindegren et al. (2021a), and distances were then derived. Using these distances and the sky positions of our sources, we queried the Bayestar19 three-dimensional dust map (G. M. Green et al. 2019) with the `dustmaps` Python package (G. Green 2018). For each source, we retrieved the 16th, 50th, and 84th percentiles of the extinction distribution, scaled them by 0.884 to convert to $E(B-V)$[9], and adopted the median as the reference value with half the difference between the upper and lower percentiles as its uncertainty.

Bayestar19 is constructed from photometry of the Gaia mission (Gaia Collaboration et al. 2018), the Panoramic Survey Telescope and Rapid Response System (Pan-STARRS1; K. C. Chambers et al. 2016), and the Two Micron All Sky Survey (2MASS; M. F. Skrutskie et al. 2006), covering wavelengths from 4000–24,000 Å (G. M. Green et al. 2019). Its spatial coverage extends across the footprint of the LAMOST survey (decl. $>-10°$). In regions with sparse coverage or missing data, Bayestar19 sometimes returns `NaN` values or zero extinction, which may be unreliable. In our parent sample of more than 4 million AFGK-type stars, approximately 16% of the sources have Bayestar19 extinction values that are either zero or undefined. In such cases, we adopted extinction estimates from the two-dimensional dust map of D. J. Schlegel et al. (1998; SFD) as a reference. We assumed a standard average extinction law for the diffuse interstellar medium with $R(V) = 3.1$. All dereddening was performed using the reddening curve from the Fitzpatrick parameterization (E. L. Fitzpatrick 1999), which is valid from the infrared to the far-ultraviolet (FUV).

After correcting for extinction, we identified sources located along the MS in the Gaia optical CMD, as shown in Figure 1. The empirical MS locus was traced using a reference sample of nearby (<100 pc) Gaia DR3 stars following the method of I. Pelisoli & J. Vos (2019). Based on this reference, we defined a relatively broad polygonal region that encloses the MS regime, explicitly accounting for uncertainties introduced by extinction corrections, different dust maps, extinction curves, and photometric errors. The final reliable WDMS candidates obtained after all selection steps are plotted as orange star symbols. Their intrinsic colors, $(B_p - R_p)_0 \approx 0.4\text{–}1.5$ mag, with 82% of the candidates having $(B_p - R_p)_0 < 1$ mag, indicate that most systems host MS companions earlier than K-type.

We then applied additional photometric and astrometric quality cuts to the AFGK-type sources located within the MS region. These cuts are based on parallax and flux measurement uncertainties in the $G$, $G_{\rm BP}$, and $G_{\rm RP}$ bands. Only sources meeting the following criteria were retained:

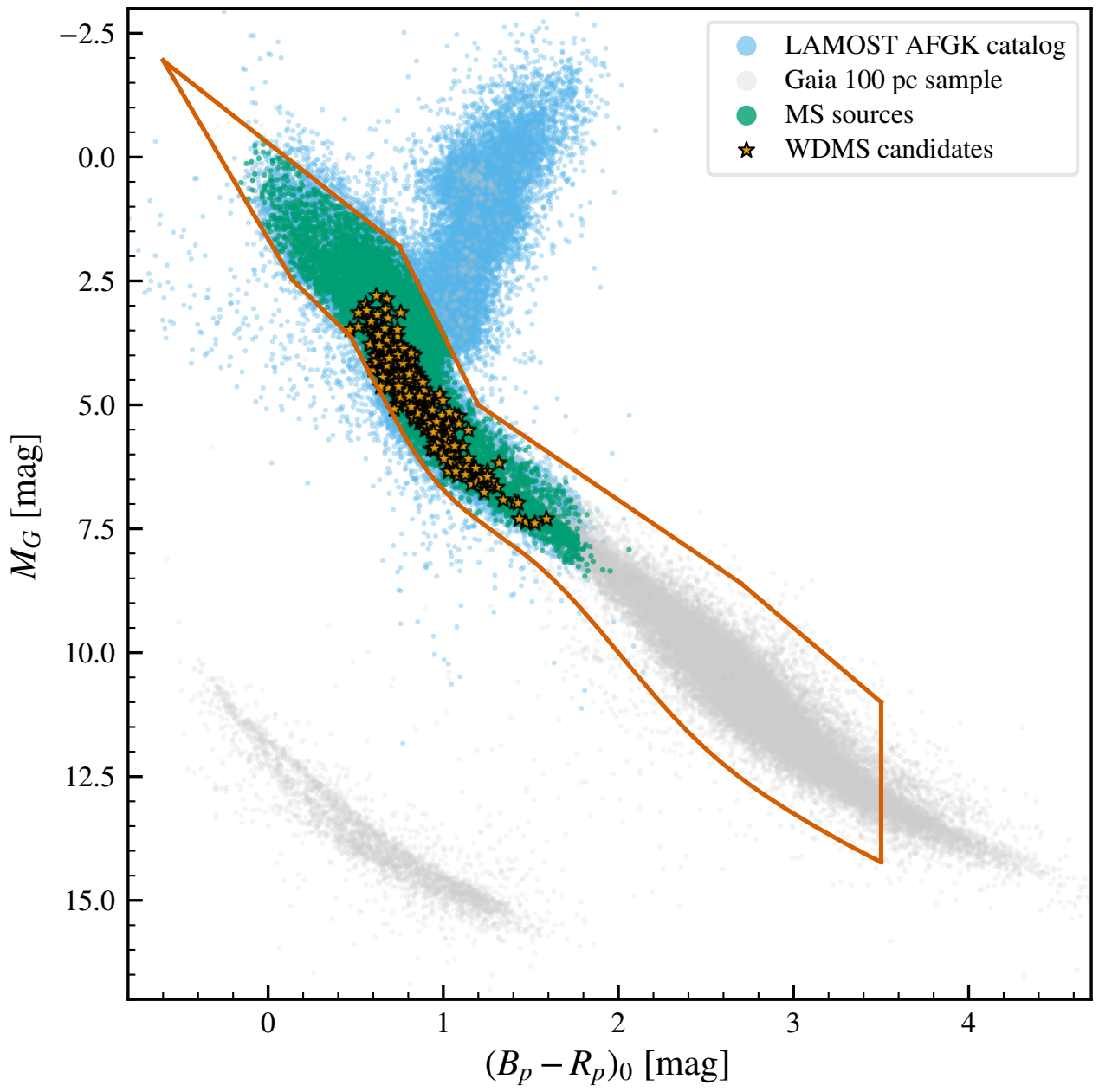

**Figure 1.** Optical color–magnitude diagram (CMD) in the absolute magnitude plane, constructed using extinction- and distance-corrected photometry from Gaia DR3. Gray points represent a reference sample of stars within 100 pc, following I. Pelisoli & J. Vos (2019), and are used to trace the empirical main-sequence (MS) locus. The red polygon delineates the adopted MS selection region. Blue points show LAMOST AFGK-type sources after the signal-to-noise and Gaia crossmatch requirements (Step 2 in Table 2). Green points represent sources that remain after applying the MS region and Gaia quality selections (Step 6 in Table 2). Orange star symbols denote the final WDMS binary candidates identified in this work after all selection steps.

1. $\varpi/\sigma_{\varpi} \geqslant 10$,
2. $F_G/\sigma_{F_G} \geqslant 10$,
3. $F_{\rm BP}/\sigma_{F_{\rm BP}} \geqslant 10$,
4. $F_{\rm RP}/\sigma_{F_{\rm RP}} \geqslant 10$,

with $\varpi$ denoting the parallax, and $F_G$, $F_{\rm BP}$, and $F_{\rm RP}$ denoting the fluxes in the corresponding bands. The $\sigma$ values are the associated standard errors. These criteria ensure that only sources with reliable astrometric and photometric data are included for further analysis.

To reduce contamination from nearby sources or bright backgrounds affecting the $G_{\rm BP}$ and $G_{\rm RP}$ bands, we applied an additional quality filter based on the Gaia photometric excess factor (D. W. Evans et al. 2018). Following M. Riello et al. (2021), we adopted the corrected excess factor $C = (F_{\rm BP} + F_{\rm RP})/F_G$ −QJ;$f(G_{\rm BP} - G_{\rm RP})$, where $f(G_{\rm BP} - G_{\rm RP})$ is a color-dependent empirical function. Values of $C^*$ close to zero indicate reliable photometry. We required $C^*$ to be smaller than the $1\sigma$ scatter defined by Equation (18) of M. Riello et al. (2021), which characterizes high-quality Gaia photometry.

We further required the renormalized unit weight error (RUWE) to be <2. Although Gaia recommends RUWE < 1.4 for reliable astrometry, higher values may also indicate unresolved binarity (L. Lindegren et al. 2012). Adopting this relaxed threshold allows us to retain potentially interesting systems while ensuring data quality. After applying this criterion, the sample was reduced to 2,342,395 sources, shown as green points in Figure 1.

[9] http://argonaut.skymaps.info/usage#units

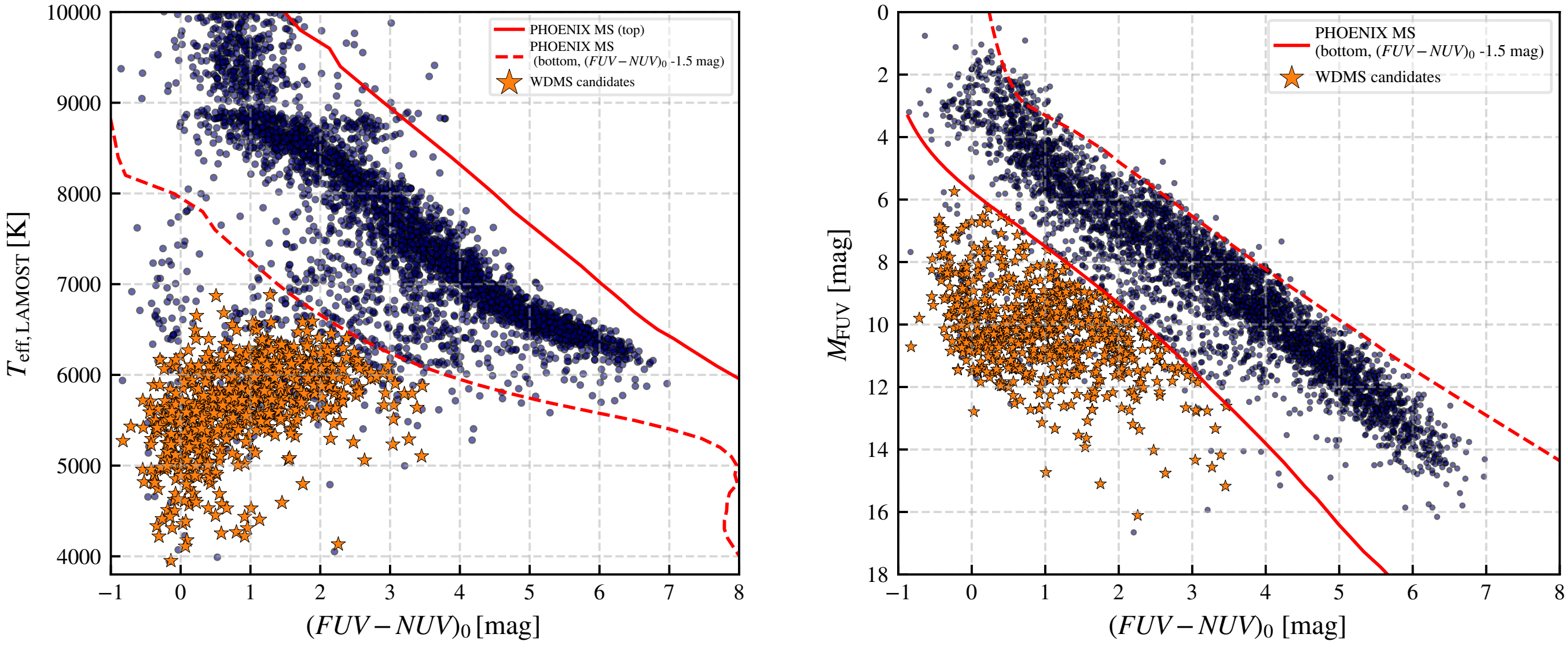

**Figure 2.** Ultraviolet color–temperature and CMDs for 15,320 extinction-corrected sources with reliable Gaia and GALEX data. Left panel: intrinsic UV color $(\mathrm{FUV}-\mathrm{NUV})_0$ vs. LAMOST spectroscopic $T_{\rm eff}$. Blue points mark sources that passed all prior quality cuts (Step 8 in Table 2). Overplotted are PHOENIX stellar atmosphere models (T. O. Husser et al. 2013): solid and dashed red curves correspond to $\log Z=+1$, $\log g=5.0$ and $\log Z=-3$, $\log g=3.5$, respectively; the dashed curve is shifted by 1.5 mag to define the UV-excess threshold. Right panel: absolute FUV magnitude vs. intrinsic $(\mathrm{FUV}-\mathrm{NUV})_0$. The same PHOENIX tracks are shown, with the solid curve shifted by 1.5 mag. Orange star symbols indicate the 654 WDMS candidates identified as UV-excess sources in this work.

### 2.3. Identification of White-dwarf Components

WD binaries with AFGK-type companions are difficult to identify from optical data alone because the luminous MS star generally dominates the flux (A. Rebassa-Mansergas et al. 2010). Motivated by the fact that a hot WD can outshine its companion in the UV even when it is invisible in the optical, we searched for systems that lie along the MS in the optical CMD but display an excess of UV flux. To identify such systems, we crossmatched the MS subsample selected in Section 2.2 with UV photometry from the Galaxy Evolution Explorer (GALEX) to reveal the WD contribution.

The GALEX mission was a NASA UV space telescope that surveyed the sky in the NUV and FUV bands (D. C. Martin et al. 2005). We crossmatched the current sample with the GALEX GR6+7 data release (L. Bianchi et al. 2017). To ensure reliable UV measurements, we applied the following criteria:

1. $e_{\rm NUV}\leqslant 0.2$ and $e_{\rm FUV}\leqslant 0.2$,
2. `NUV_art` $=0$ and `FUV_art` $=0$,
3. `NUV_ext` $=0$ and `FUV_ext` $=0$.

Here, $e_{\rm NUV}$ and $e_{\rm FUV}$ denote photometric uncertainties in the NUV and FUV bands. The flags `NUV_art` and `FUV_art` identify contamination from image artifacts, while `NUV_ext` and `FUV_ext` mark extended or blended sources. All objects with nonzero values in any of these flags were removed to avoid unreliable or poorly extracted photometry.

After correcting for extinction, we examined two diagnostic relations for sources that passed all prior quality cuts, shown as blue points in Figure 2. The left panel shows the intrinsic UV color $(\mathrm{FUV}-\mathrm{NUV})_0$ as a function of the LAMOST spectroscopic temperature $T_{\rm eff,LAMOST}$, while the right panel plots $(\mathrm{FUV}-\mathrm{NUV})_0$ against the absolute FUV magnitude $M_{\rm FUV}$. PHOENIX stellar atmosphere models (T. O. Husser et al. 2013) are overplotted as red curves, representing the expected UV colors of single MS stars with different metallicities and surface gravities (S. G. Parsons et al. 2016; A. Rebassa-Mansergas et al. 2017). To identify UV-excess sources, the lower model sequence in each panel is shifted by 1.5 mag toward bluer colors, defining the boundary below which objects are likely to host hot WD companions. The robust WDMS candidates are highlighted as orange star symbols in both panels.

Guided by the distributions in Figure 2, we classified a source as showing UV excess if its intrinsic color $(\mathrm{FUV}-\mathrm{NUV})_0$ lies at least 1.5 mag below the adopted lower MS boundary, defined by the blue-shifted PHOENIX track applied in both panels. To minimize contamination from non-WD binaries, we further restricted the sample to $(\mathrm{FUV}-\mathrm{NUV})_0<3.5$, as systems with redder colors are likely dominated by nondegenerate companions (B. Anguiano et al. 2022). Applying these criteria yields 772 WDMS binary candidates exhibiting clear UV excess. In these systems, the WD dominates the UV flux while the MS companion dominates the optical, causing the binaries to overlap with the MS locus in the optical CMD (Figure 1) but shift toward the WD sequence in the UV CMD—a characteristic signature of unresolved WDMS binaries.

### 2.4. Spectral Energy Distribution Fitting

To further characterize the stellar parameters of the 772 candidates, we performed binary spectral energy distribution (SED) fitting using multiband photometric data. For this purpose, we modified the publicly available `ARIADNE` package developed by J. I. Vines & J. S. Jenkins (2022), which was originally designed for single-star SED fitting. The single-component flux model was replaced with the sum of two model spectra, one for the MS star and the other for the WD, to enable fitting of unresolved binaries. The combined theoretical SED was then compared directly with the observed photometric fluxes.

For the MS component, we adopted the PHOENIX stellar atmosphere models (T. O. Husser et al. 2013), which span effective temperatures from 2300–120,00 K, surface gravities in the range $0.0 \leqslant \log g \leqslant 6.0$, and metallicities between $-4.0$ and $+1.0$, without accounting for $\alpha$-element enhancement or depletion. For the WD component, we used the hydrogen-rich (DA) model grid from D. Koester (2010), covering $5000 \leqslant T_{\rm eff} \leqslant 80{,}000$ K with $6.5 \leqslant \log g \leqslant 9.5$. The temperature grid is nonuniform, with step sizes of 250–1000 K depending on temperature.

The binary SED fitting involves nine free parameters: the effective temperatures, surface gravities, and radii of both stellar components, the metallicity of the MS star, the system distance, and the extinction $A_V$. Gaussian priors are adopted for the MS effective temperature and distance, centered on the LAMOST spectroscopic values and the Gaia-based distance, respectively. For extinction, we adopt Gaussian priors based on the Bayestar19 estimates. When Bayestar19 returns a value of zero or NaN, we instead adopt a uniform prior ranging from $A_V = 0$ to the corresponding SFD extinction value. This applies to approximately 27% of the systems in the SED fitting stage. This treatment ensures that extinction values close to zero remain accessible during the sampling process, so that if the true extinction is negligible, the SED fitting naturally converges toward $A_V \approx 0$. For the WD, the surface gravity is assigned a uniform prior over $6.5 \leqslant \log g \leqslant 9.5$, and the radius is sampled uniformly between 0.0001 and $0.5\,R_\odot$. All other parameters adopt the default priors provided by ARIADNE. While surface gravity and metallicity are weakly constrained by SEDs, they are included to improve the fit stability.

The likelihood function is assumed to be Gaussian, following the formulation of J. I. Vines & J. S. Jenkins (2022):

$$\mathcal{L} = \prod_{i=1}^{N} \frac{1}{\sqrt{2\pi\epsilon_i^2}} \exp\left(-\frac{1}{2}\left(\frac{f_i - {\rm SF}_{i,M}}{\epsilon_i}\right)^2\right),$$

where $f_i$ is the observed flux in the $i$th photometric band, ${\rm SF}_{i,M}$ is the corresponding synthetic flux from the combined MS +WD model, and $\epsilon_i$ denotes the total uncertainty, obtained by adding in quadrature the measurement error $\sigma_{i,\rm m}$ and the excess noise $\sigma_{i,\rm e}$, i.e., $\epsilon_i^2 = \sigma_{i,\rm m}^2 + \sigma_{i,\rm e}^2$ (see J. I. Vines & J. S. Jenkins 2022). The fitting incorporates photometric data from GALEX, Gaia, 2MASS, and All WISEW1 and W2 (E. L. Wright et al. 2010), with additional measurements from the Sloan Digital Sky Survey (D. J. Eisenstein et al. 2011), Pan-STARRS1, and APASS (A. A. Henden et al. 2015) when available. Data points without reported uncertainties or with errors larger than 0.2 mag were excluded.

After performing the binary SED fitting with a nested sampling algorithm, we obtained posterior distributions for all model parameters and derived best-fit values. Based on these fits, our modified version of ARIADNE outputs the visual goodness-of-fit parameter ($Vgf_{\rm b}$),[10] as defined in the Virtual Observatory SED Analyzer (VOSA; A. Bayo et al. 2008), which we use as a reference indicator of fit quality.

A source is considered reliably fitted if it satisfies two criteria. First, the $Vgf_{\rm b}$ value must be less than or equal to 15. Second, the absolute fractional residual flux, defined as

$$f_{\rm residual} = \left| \frac{F_{\lambda,\rm obs} - F_{\lambda,\rm model}}{F_{\lambda,\rm obs}} \right|,$$

must individually satisfy $f_{\rm residual} < 0.5$ in both UV bands (FUV and NUV), with at least 80% of the flux points in the optical-infrared regime also meeting this condition (P. K. Nayak et al. 2024). Applying these filters, we finally identified 759 WDMS candidates with well-constrained SED fits.

Well-fitted WDMS candidates are shown in Figure 3. For clarity, the panels are arranged such that systems within the same column have similar WD effective temperatures, while those within the same row have comparable MS temperatures. These examples illustrate the diversity of the WDMS sample across a broad range of stellar parameters. In each system, the MS component dominates the optical flux, whereas the WD contributes primarily at UV wavelengths. This wavelength-dependent flux contrast explains why these systems follow the MS locus in the optical CMD but appear significantly bluer in the UV CMD.

For a small subset of sources, reliable SED fits could not be obtained. The failures were mainly due to insufficient photometric coverage or to inconsistencies among fluxes from different surveys, which prevented a coherent two-component fit. These cases reflect the typical limitations of the fitting procedure rather than genuine astrophysical anomalies. Table A1 shows a representative subset of the 772 UV-excess WDMS candidates, and Table A2 lists a subset of the 759 well-fitted systems. The complete versions of both tables are available.

## 3. Refinement of the WDMS Candidate Sample

In this section, we refine the WDMS candidate sample by identifying and flagging potential sources of contamination to obtain a cleaner and more reliable catalog, corresponding to Steps 11–13 in Table 2.

### 3.1. Chromospheric Activity in Main-sequence Stars

As described in Section 2.3, the identification of WD components in our WDMS candidates relies on UV excess, which generally indicates the presence of a hot WD. However, chromospherically active MS stars can also produce enhanced UV emission. In single stars, such activity originates from magnetic heating in their outer atmospheres, and in close binaries, it can be further strengthened by effects such as irradiation or weak accretion (G. H. Smith 2019; L.-y. Zhang et al. 2023; A. Shi et al. 2025). These active stars may exhibit unusually blue UV colors and fall into the FUV-bright region of the UV CMD. Although genuine WDMS systems can also show magnetic activity through binary interaction, strong activity without additional evidence of a WD companion suggests that the UV excess may arise from stellar activity rather than from a hot WD.

Chromospherically active stars are also expected to have X-ray counterparts. To assess this possibility and ensure a cleaner WDMS sample, we crossmatched our 759 well-fitted WDMS candidates with both X-ray and optical catalogs of stellar activity. The X-ray comparison included the Chandra Source Catalog (I. N. Evans et al. 2010), XMM-Newton (R. D. Saxton et al. 2008; N. A. Webb et al. 2020), the ROSAT All-Sky Survey (T. Boller et al. 2016), and the eROSITA All-Sky Survey (A. Merloni et al. 2024). For optical activity samples, we used R. Martínez-Arnáiz et al. (2010), S. Boro Saikia et al. (2018), C. E. Brasseur et al. (2019), and J. Gomes da Silva et al. (2021). No counterparts were found in the optical activity catalogs, and only 15 matches were identified in the X-ray surveys. The small number of X-ray detections suggests that contamination from chromospherically active stars is likely minor.

[10] https://svo2.cab.inta-csic.es/theory/vosa/helpw4.php?otype=star&action=help&what=fit

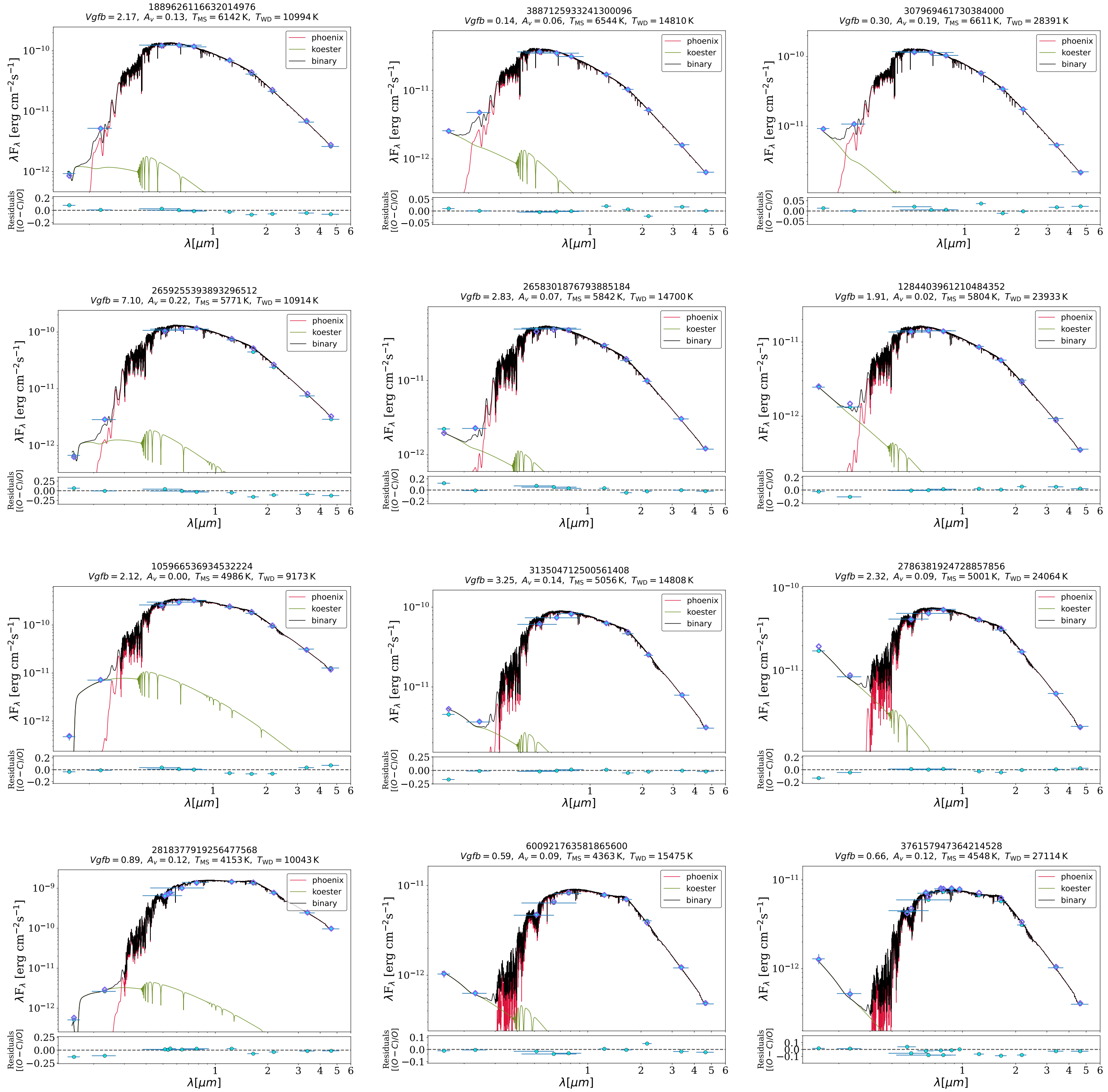

**Figure 3.** Best-fit binary spectral energy distributions for representative WDMS candidates. The plot titles list the source identifier and fitted parameters: Gaia source ID, $Vgf_b$, $A_V$, $T_{MS}$, and $T_{WD}$. Here, $Vgf_b$ is the visual goodness-of-fit parameter, defined as a modified reduced $\chi^2$; smaller values indicate better fits. The red and green curves represent the fitted spectra of the MS star (PHOENIX models) and the WD (Koester DA models), respectively, while the black curve shows their combined spectrum. Observed photometric fluxes are shown as blue circles, with horizontal bars indicating the wavelength ranges of the corresponding filters; magenta diamonds denote the model-predicted fluxes. The lower panels display the fractional residuals $(O - C)/O$ for each band. The displayed systems span a range of MS and WD effective temperatures.

Like V471 Tau, a well-known active WDMS binary (G. A. J. Hussain et al. 2006), chromospheric activity can also be present in the MS stars of WDMS systems. To assess its potential impact on our WDMS candidates, we computed the $R^+$ index from the Ca II H and K lines, which are more sensitive to chromospheric emission than the Ca II infrared triplet (J. Martin et al. 2017). For each available LAMOST spectrum, we derived the $R^+$ values for the H and K lines separately and adopted their average, $R^+_{\rm HK}$, as the final activity indicator. For sources with multiple spectra, the mean $R^+_{\rm HK}$ was used as the representative value.

Figure 4 shows the chromospheric activity index $R^+_{\rm HK}$ as a function of the MS effective temperature for all 759 WDMS candidates. The activity increases toward cooler stars, as expected for late-type dwarfs. Most candidates have G-type primaries with modest activity levels: 85% have $R^+_{\rm HK} < 0.2$, and 53% fall below 0.1, indicating that the sample is not dominated by chromospheric emission. One notable outlier is a highly active system with $R^+_{\rm HK} = 0.99$, corresponding to a K7 dwarf. Gaia DR3 reports 20 radial-velocity (RV) transits with an amplitude of 100.46 km s$^{-1}$, and the source is classified as a BY Dra variable, a typical signature of magnetically active binaries. Although such variability may arise from single active stars, it can also be induced by binarity (B. W. Bopp & F. Fekel 1977; Z. Eker et al. 2008), so this system remains a plausible WDMS candidate pending further confirmation.

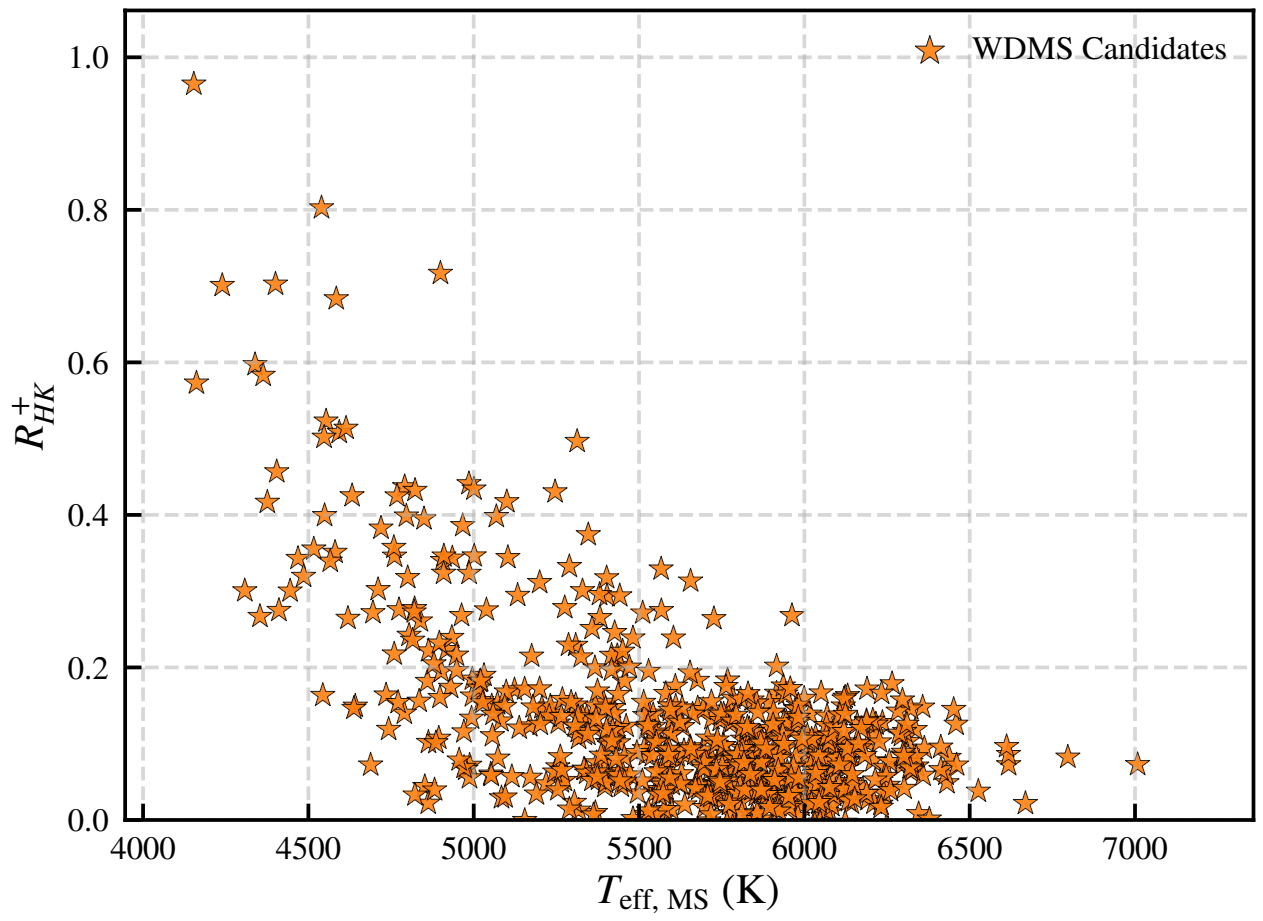

**Figure 4.** Chromospheric activity in our WDMS candidates, shown as $R^+_{\rm HK}$ vs. the MS effective temperature derived from SED fitting. The index $R^+_{\rm HK}$ is measured from LAMOST low-resolution spectra following the method of X. Huang et al. (2024). Each star symbol corresponds to one candidate. The majority of systems are concentrated in the G-type range and display low activity, suggesting that contamination from active stars is limited.

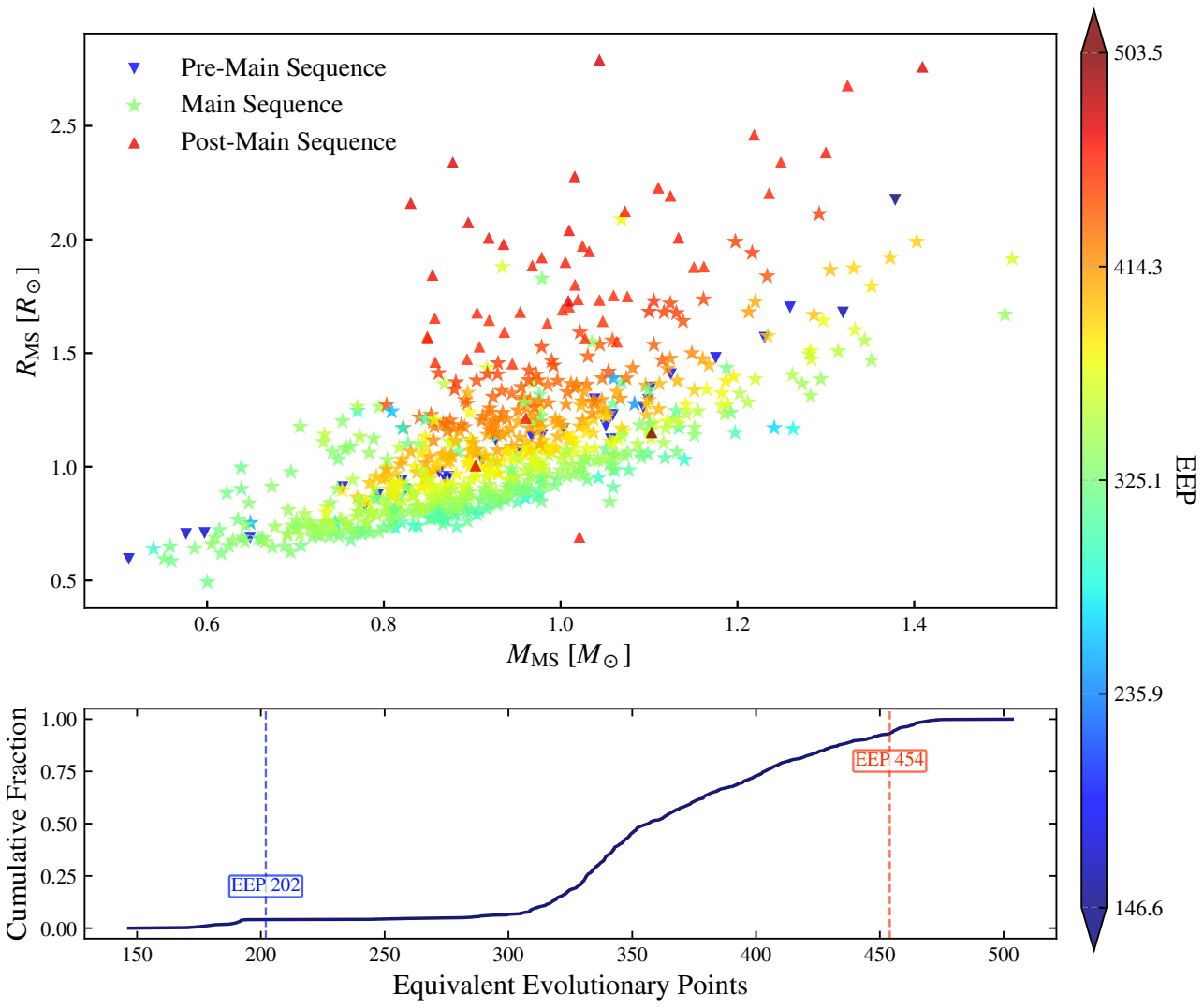

**Figure 5.** Mass–radius distribution and evolutionary status of the MS companions in the WDMS candidates. Top panel: distribution in the mass–radius plane, color coded by evolutionary stage inferred from equivalent evolutionary point (EEP) values. Star symbols represent MS companions ($202 \leqslant \rm EEP \leqslant 454$), blue inverted triangles mark possible pre-MS stars ($\rm EEP < 202$), and red upright triangles denote potential post-MS stars ($\rm EEP > 454$). Most companions have masses between 0.8 and 1.2 $M_\odot$; those with larger radii are mainly associated with post-MS evolution. Bottom panel: cumulative distribution of EEP values. Vertical dashed lines indicate the zero-age main sequence (ZAMS; EEP = 202) and terminal-age main sequence (TAMS; EEP = 454). Both pre-MS and post-MS stars appear to represent only a small fraction of the sample, while the median EEP of 356 suggests that most companions are in the mid-to-late MS phase.

Previous studies have investigated the correlation between chromospheric activity indicators derived from the Ca II H and K lines and X-ray emission, generally finding a positive relationship (C. J. Schrijver 1983; M. Mittag et al. 2018; B. Fuhrmeister et al. 2022).

Among the 15 systems in our sample with X-ray detections, the median and mean values of $R^+_{\rm HK}$ are 0.23 and 0.28, respectively. In contrast, the remaining sources exhibit significantly lower activity levels, with a median of 0.09 and a mean of 0.12. Within the full sample of 759 systems, 7.7% of sources with $R^+_{\rm HK} > 0.2$ have X-ray detections, whereas only 0.79% of those with $R^+_{\rm HK} \leqslant 0.2$ are detected in X-rays. These results consistently indicate that systems with larger $R^+_{\rm HK}$ values are more likely to exhibit X-ray emission.

We therefore do not remove objects solely on the basis of their $R^+_{\rm HK}$ values. Instead, we report the activity indices for all candidates and flag the X-ray detections identified from external catalogs, as listed in Table A2, to facilitate future assessments of possible contamination by active stars.

### *3.2. Misclassified Non-MS Companions*

Although MS companions were selected based on their positions in the Gaia CMD (Figure 1), a small number of pre-MS or subgiant stars may still overlap with the MS locus. Such stars, either approaching the zero-age main sequence (ZAMS) or evolving away from it, have similar photometric properties and can therefore be misclassified as MS companions within our WDMS candidate sample.

We evaluated the evolutionary status of the MS companions by estimating their physical properties with the latest version of the `isochrones` package (T. D. Morton 2015), which interpolates the MIST (J. Choi et al. 2016; A. Dotter 2016) using `MultiNest` (F. Feroz et al. 2009; J. Buchner et al. 2014). For each WDMS candidate, the inputs include the metallicity from LAMOST spectroscopy, the Gaia $G$, $G_{\rm BP}$, and $G_{\rm RP}$ magnitudes with their uncertainties, together with the SED-derived parameters of the MS star. The resulting posterior distributions of stellar mass, age, and equivalent evolutionary points (EEPs) serve as indicators of evolutionary stage.

As defined by A. Dotter (2016), EEPs denote key stages along stellar evolutionary tracks and allow interpolation between models with different initial masses. Following the classification of M. G. Soto et al. (2021), stars with EEP < 202 are considered pre-MS, those with $202 \leqslant \rm EEP \leqslant 454$ are classified as MS, and those with EEP > 454 are identified as post-MS, including subgiant and RGB phases.

Adopting this classification, Figure 5 shows the distribution of WDMS candidates in the mass–radius plane, color coded by the evolutionary stage of their MS companions. Most systems have companions with masses between 0.8 and 1.2 $M_\odot$. This approach avoids relying on LAMOST low-resolution $\log g$ measurements for mass determination, as those surface-gravity estimates often have large uncertainties and may be biased or capped (e.g., $\Delta \log g \sim 0.9$ dex). Sources classified as post-MS exhibit systematically larger radii, consistent with subgiant evolution, whereas the pre-MS group constitutes only a small fraction. The cumulative distribution of EEP values indicates that most companions occupy the mid-to-late MS phase, with a median EEP = 356. Because the evolutionary classification depends on model interpolation and the precision of the input

**Table 3**
Wide WDMS Binary Candidates Identified within 6″

| Query Coord. | Gaia DR3 Source ID | $\varpi \pm \sigma_{\varpi}$ (mas) | $\mu_{\alpha} \pm \sigma_{\mu_{\alpha}}$ (mas $yr^{-1}$) | $\mu_{\delta} \pm \sigma_{\mu_{\delta}}$ (mas $yr^{-1}$) | $G$ (mag) | Sep. (arcsec) | Projected Sep. (au) | $R$ |
|---|---|---|---|---|---|---|---|---|
| 42.303, 9.327 | 20808983405832960 | 1.23 ± 0.02 | 4.27 ± 0.03 | −8.08 ± 0.03 | 14.27 | 5.48 | 4443 | 0.0136 |
| | 20808983405224064 | 1.13 ± 0.46 | 5.40 ± 0.59 | −8.31 ± 0.53 | 19.62 | ... | ... | ... |
| 121.335, 21.594 | 670964330763140224 | 1.89 ± 0.03 | 17.97 ± 0.04 | −16.90 ± 0.02 | 14.83 | 2.62 | 1387 | 0.0003 |
| | 670964330763660800 | 2.14 ± 0.48 | 17.20 ± 0.54 | −16.92 ± 0.33 | 19.88 | ... | ... | ... |
| 145.432, 26.271 | 646674602780526848 | 1.19 ± 0.03 | −11.51 ± 0.04 | −4.44 ± 0.03 | 13.96 | 4.98 | 4179 | 0.0015 |
| | 646674671501401984 | 1.49 ± 0.44 | −11.09 ± 0.44 | −4.47 ± 0.38 | 19.68 | ... | ... | ... |
| 251.228, 15.052 | 4461890379624760960 | 1.57 ± 0.03 | −7.47 ± 0.03 | −1.78 ± 0.02 | 15.25 | 2.61 | 1665 | 0.0001 |
| | 4461890379624760832 | 1.67 ± 0.22 | −7.31 ± 0.29 | −1.52 ± 0.16 | 18.76 | ... | ... | ... |

**Note.** Parameters are provided for both components.

parameters, some systems may be misclassified, but all remain plausible WDMS candidates.

For the 759 WDMS candidates with well-fitted SEDs, the median stellar masses and ages from the posterior distributions are provided in Table A2, where we also flag whether the MS companion is consistent with the main-sequence evolutionary phase.

### 3.3. Wide WDMS Systems and Chance Alignments

Chance alignments between unrelated MS and WD stars can also lead to false positives in our WDMS sample. Because GALEX has a relatively low angular resolution of about 6–8″,[11] a wide WDMS pair may appear as a single UV source. In contrast, Gaia achieves an angular resolution of 0.″4–0.″5 (Gaia Collaboration et al. 2018, 2021) and can resolve individual components at such separations. As a result, in wide WDMS systems, Gaia may detect the WD and MS components as separate sources. When crossmatched with the lower-angular-resolution GALEX data, the UV emission may be associated only with the isolated WD, thereby mimicking the signature of an unresolved WDMS binary.

To investigate this possibility, we crossmatched our 759 WDMS candidates with the catalog of WD candidates in Gaia EDR3 from N. P. Gentile Fusillo et al. (2021) and identified nine spatially coincident sources. For each case, we searched within a radius of 6″ around the corresponding Gaia DR3 position to retrieve nearby stellar counterparts and assess whether they form physically associated systems. Following the binary-association criteria adopted by K. El-Badry et al. (2021), we classified two stars as a neighboring pair if they satisfy:

1. a projected separation smaller than 5 pc,
2. a proper-motion difference <5 km $s^{-1}$,
3. parallaxes consistent within $2\sigma$.

Applying these criteria, four of the nine matched cases are identified as likely wide WDMS binaries, each consisting only of a mutual neighboring pair. The remaining systems are consistent with chance alignments. Table 3 lists the four wide candidates together with their Gaia DR3 source IDs, parallaxes, proper motions, $G$-band magnitudes, angular separations, projected physical separations, and the corresponding chance-alignment probabilities.

Among these four systems, the first three do not appear in the final wide-binary catalog of K. El-Badry et al. (2021) because their WD components have $\varpi/\sigma_{\varpi} < 5$, and were therefore excluded in that work. Nevertheless, following the method described in K. El-Badry et al. (2021), we estimated the $R$ values for all four systems as a proxy for the chance-alignment probability. All four systems have $R < 0.1$, indicating a low probability of random alignment. We therefore consider them high-probability wide WDMS binaries and include them in our sample, while noting that further observational confirmation is still required.

Only one WD candidate, Gaia DR3 646674671501401984, has a published mass estimate in N. P. Gentile Fusillo et al. (2021), derived using pure-hydrogen atmosphere models. Their value of 0.60 $M_{\odot}$ agrees well with our SED-based estimate of 0.61 $M_{\odot}$, offering useful validation of our method (see Section 4.2). This WD is paired with the MS star Gaia DR3 646674602780526848 at an angular separation of 4.98″, corresponding to a projected physical distance of 4179 au. Assuming a circular orbit, Kepler's third law gives an orbital period of $\sim 2.2 \times 10^{5}$ yr, indicating that the two stars have evolved independently. The MS companion shows an RUWE value of 1.68 in Gaia DR3, exceeding the commonly used threshold of 1.4 (L. Lindegren et al. 2021b). Because the distant WD cannot induce such astrometric deviations, the elevated RUWE may point to an unresolved close companion (V. Belokurov et al. 2020; A. Castro-Ginard et al. 2024). LAMOST spectra obtained in 2011 and 2024 provide six single-epoch RV measurements for the MS star, which show no significant variations beyond the instrumental uncertainties (Z.-R. Bai et al. 2021). Thus, the presence of a potential inner companion cannot be confirmed with the current data.

The remaining Gaia sources show inconsistent parallaxes or large proper-motion differences and are therefore not gravitationally bound. These cases are flagged accordingly in Table A2.

### 3.4. Potential Contamination from Hot Subdwarfs

Our identification of WD components relies on UV excess, which generally indicates the presence of a hot WD. However, some compact UV-bright objects, particularly hot-subdwarf stars, can also produce strong UV emission and may contaminate a UV-based WDMS selection. Hot subdwarfs are core helium-burning stars that have lost their hydrogen envelopes and typically have masses of about 0.47 $M_{\odot}$

[11] https://asd.gsfc.nasa.gov/archive/galex/Documents/instrument_summary.html

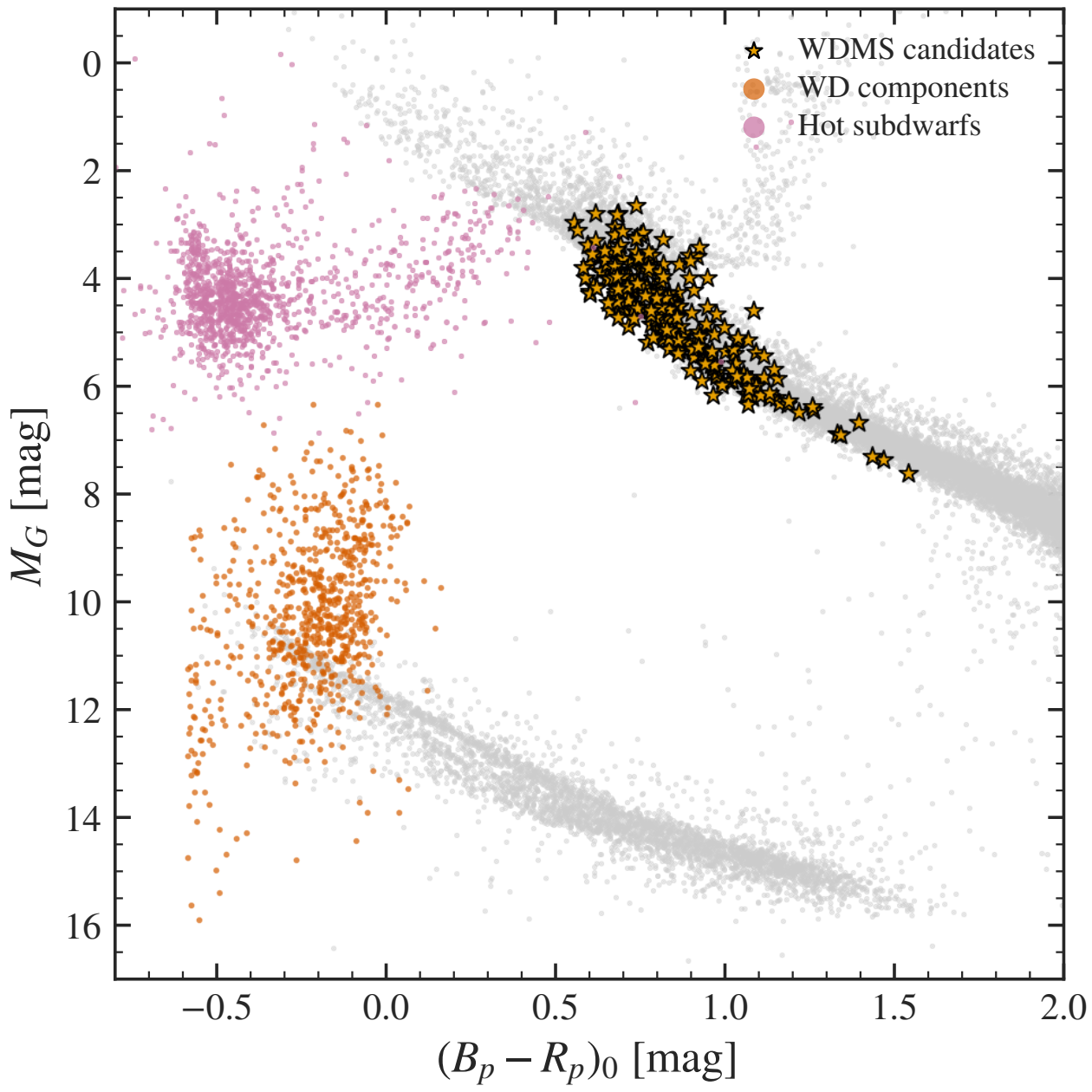

**Figure 6.** Similar to Figure 1, this optical CMD shows the synthesized Gaia magnitudes of the WD components (orange points), computed from the SED-derived WD model spectra. These predicted WD loci are compared with the hot-subdwarf sample from S. Geier (2020) and Z. Lei et al. (2021) (pink points).

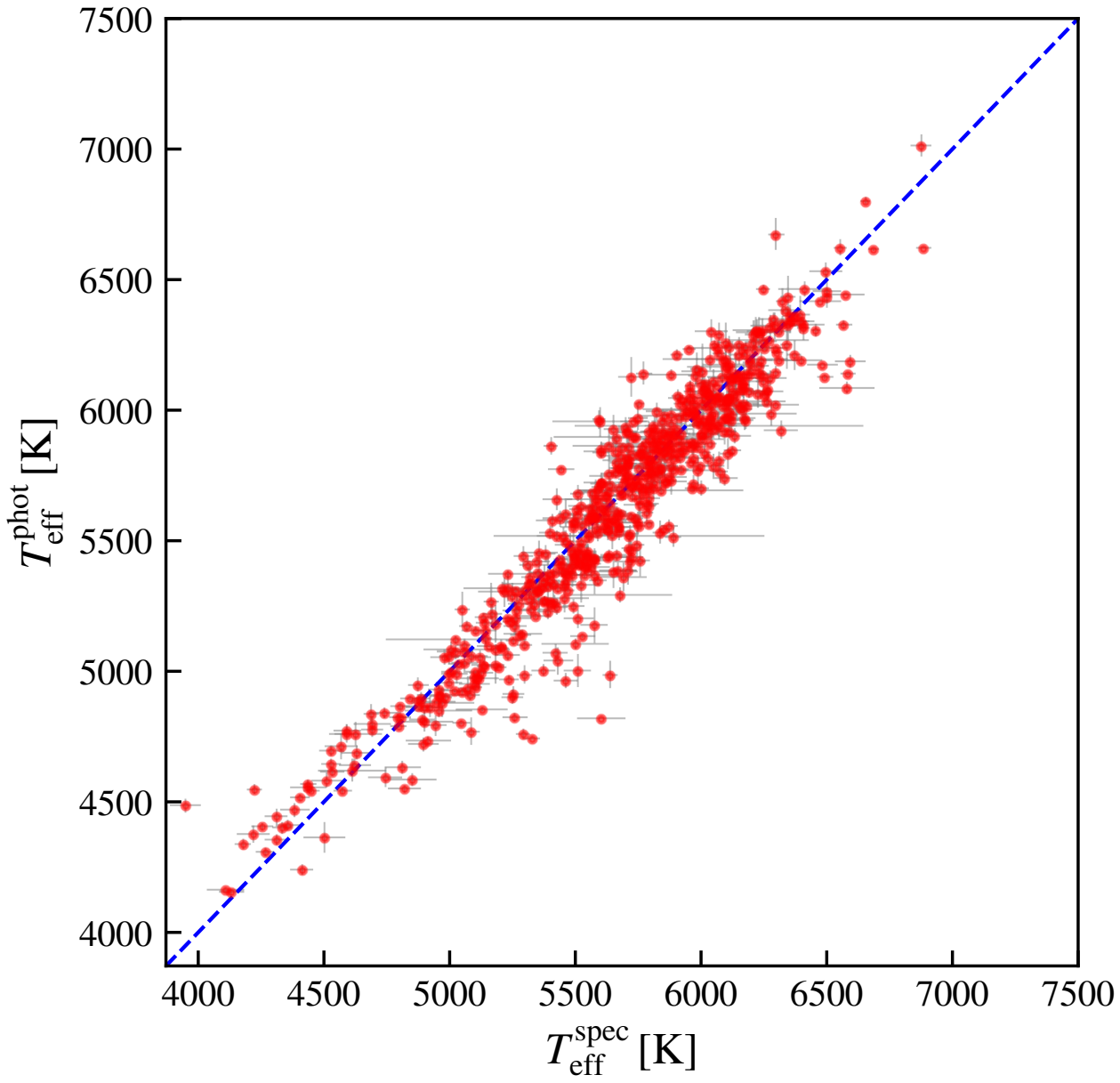

**Figure 7.** Comparison between the spectroscopic and SED-derived effective temperatures for the MS components in our WDMS candidates. Each red point represents one system, with $T_{\rm eff}^{\rm spec}$ from LAMOST spectroscopy and $T_{\rm eff}^{\rm phot}$ from SED fitting. Overall, the two estimates show good consistency, indicating reliable temperature determinations for the MS companions.

(U. Heber 2024). They are bright in the UV, similar to ELM WDs, but remain substantially more luminous in the blue optical bands. For this reason, the color cut that removes sources bluer than the MS region in the Gaia CMD (Figure 1) should exclude hot subdwarfs when paired with F-, G-, or K-type companions. Nonetheless, as shown in Figure 5, some companions may be more evolved and therefore more luminous, and such systems might not be removed by this color-based criterion alone.

To examine this possibility, Figure 6 compares the predicted positions of the WD components, computed from our SED-derived WD models, with the hot-subdwarf samples of S. Geier (2020) and Z. Lei et al. (2021). The predicted WD locus is clearly separated from the hot-subdwarf sequence, indicating that contamination from hot subdwarfs is negligible.

The diagram also shows that many WD components occupy relatively bright absolute magnitudes, consistent with a substantial population of low-mass WDs in our candidates.

In addition, we repeated the binary SED fitting described in Section 2.4 by replacing the DA WD model grid with atmosphere models from the Tübingen non-LTE (NLTE) Model-Atmosphere Package (TMAP; K. Werner et al. 2012). We find that adopting the TMAP models does not produce any significant change in the predicted distribution of the WD components, demonstrating that our conclusions are not sensitive to the assumption of WD model atmospheres.

We further note that J. A. Garbutt et al. (2024) proposed flagging systems with $M_{\rm FUV} < 5.2$ mag as hot-subdwarf binary candidates. None of our WDMS candidates satisfy this empirical criterion, providing an independent indication that hot-subdwarf contamination is minimal.

## 4. Discussion

Table 2 summarizes the number of sources remaining at each stage of the selection procedure described in Sections 2 and 3. After applying all selection and cleaning steps, we obtain a refined sample of 654 high-confidence WDMS candidates. In the following, we analyze the properties of this final sample and examine its observational characteristics. These systems can be readily identified using the quality flags provided in Table A2.

### *4.1. Stellar Parameter Distributions of the MS Companions*

The LAMOST spectra provide valuable constraints on the atmospheric properties of the MS components in our WDMS candidates. For systems with LAMOST coverage, we adopt the spectroscopically determined parameters of the MS star, including the effective temperature ($T_{\rm eff}$), surface gravity ($\log g$), and metallicity ([Fe/H]). Figure 7 compares the spectroscopic temperatures ($T_{\rm eff}^{\rm spec}$) with the photometric temperatures ($T_{\rm eff}^{\rm phot}$) derived from SED fitting. Both estimates span a similar range of approximately 4000–7000 K, with a standard deviation of 151 K between the two scales, indicating good overall agreement.

For the final robust sample of WDMS candidates, Figure 8 shows the normalized distributions of the stellar parameters of the MS companions. The shaded purple histograms represent the distributions of the full sample. The quantities plotted include the effective temperature ($T_{\rm MS}$) and radius ($R_{\rm MS}$) derived from the binary SED fits, the surface gravity ($\log g$) and metallicity ([Fe/H]) measured from LAMOST spectroscopy, and the mass ($M_{\rm MS}$) estimated from isochrone modeling (Section 3.2).

The effective temperatures peak near 6000 K, with most companions located in the G-type regime and a substantial fraction in the F-type range, based on the empirical scale of M. J. Pecaut & E. E. Mamajek (2013).[12] A-type companions are essentially absent, likely because they are intrinsically less common and have shorter MS lifetimes than later-type stars, which reduces their detection probability. Their higher temperatures also provide weaker UV contrast with a WD, making such binaries more difficult to identify through UV excess. The metallicities span from $[{\rm Fe/H}] \approx -1.3$ to $+0.5$, clustering near the solar value, while the surface gravities range from $\log g \approx 3.5$ to 4.8 with a peak around 4.3. The distributions of $R_{\rm MS}$ and $M_{\rm MS}$ show broad peaks near $R_{\rm MS} \approx 0.9\,R_\odot$ and $M_{\rm MS} \approx 0.9\,M_\odot$, consistent with the typical properties of solar-type dwarfs.

[12] https://www.pas.rochester.edu/~emamajek/EEM_dwarf_UBVIJHK_colors_Teff.txt

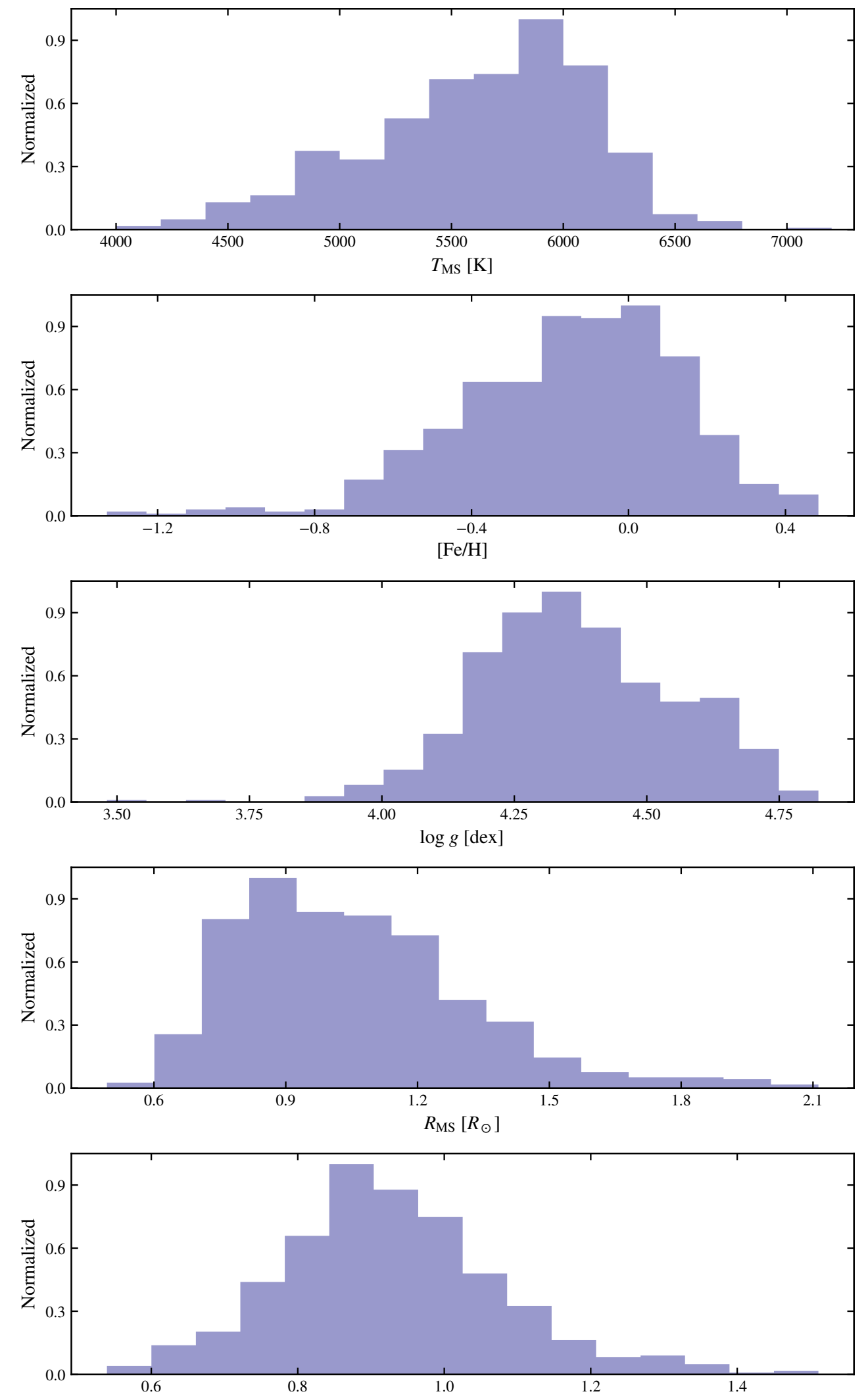

**Figure 8.** Normalized distributions of the stellar parameters of the MS companions in the refined sample of 654 WDMS candidates. The panels show the distributions of effective temperature ($T_{\rm MS}$), surface gravity ($\log g$), metallicity ([Fe/H]), radius ($R_{\rm MS}$), and mass ($M_{\rm MS}$).

### 4.2. Stellar Parameter Distributions of the WD Components

Following the analysis of the MS companions, we now examine the stellar parameters of the WD components in our WDMS candidates. Because SED fitting is largely insensitive to surface gravity, the resulting $\log g$ values cannot be used to determine WD masses. Instead, we infer the WD masses using theoretical cooling models. The effective temperatures and bolometric luminosities ($L_{\rm WD}$) obtained from our SED fits are interpolated within the Montreal white-dwarf evolutionary model grid (A. Bédard et al. 2020) using the open-source `WD_models`[13] package. In this procedure, the median values of $T_{\rm WD}$ and $L_{\rm WD}$ are matched to the nearest grid points to derive self-consistent estimates of the WD masses and cooling ages.

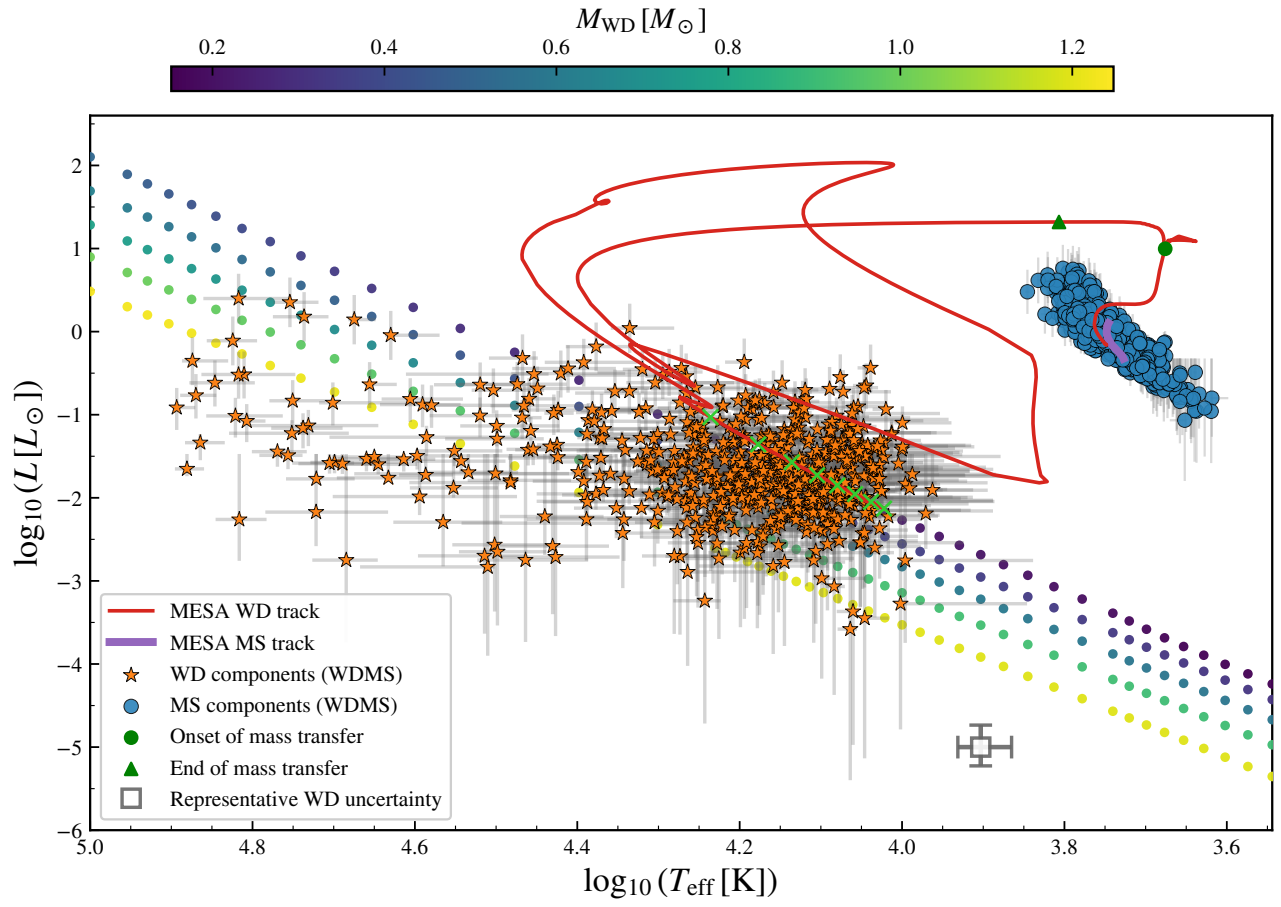

**Figure 9.** Hertzsprung–Russell diagram for the WD and MS components of the WDMS candidates, based on parameters derived from the binary SED fitting. WD components are shown as orange star symbols and MS companions as blue circles. Uncertainties in both effective temperature and luminosity are indicated by gray error bars. A representative WD uncertainty is displayed as an open gray square with error bars, shifted to the lower right for clarity. Montreal WD cooling tracks (A. Bédard et al. 2020), color coded by WD mass, are overplotted for reference. An illustrative MESA binary-evolution model is also shown to demonstrate the formation of an ELM WD; the evolutionary tracks of the progenitor and its MS companion are plotted in red and purple, respectively. A green circle marks the onset of mass transfer, a green triangle indicates its termination, and green crosses denote subsequent evolutionary stages at 300 Myr intervals after mass transfer ceases.

Figure 9 shows the distribution of WD and MS components for the WDMS binaries, based on SED-derived parameters. Measurement uncertainties in effective temperature and luminosity are shown for the WD components, with the median WD uncertainty indicated by a gray square shifted to the lower right for clarity. Most WD components cluster along cooling tracks corresponding to low-mass WDs, and a significant subset falls within the ELM WD regime.

To illustrate a representative evolutionary pathway leading to the formation of an ELM WD, we computed an example binary-evolution model using `MESA` (version r23.05.1; B. Paxton et al. 2011, 2013, 2015, 2018, 2019; A. S. Jermyn et al. 2023). The evolutionary tracks shown in Figure 9 correspond to a model with a $1\,M_\odot$ primary, a mass ratio of 1.1, and an initial orbital period of 5 days. Following A. C. Rubio et al. (2025), who showed that population-synthesis models of WD binaries are best matched with low mass-transfer efficiencies ($\beta$), we adopt $\beta = 0.2$. As the primary ascends the RGB, stable-RLOF begins (green circle) and proceeds until the donor is stripped to a $0.25\,M_\odot$ pre-ELM WD paired with a $1\,M_\odot$ MS companion (green triangle), leaving the system in a widened $\sim$13 day orbit. After mass transfer ceases, the stripped core evolves into a cooling WD while the companion continues toward the terminal-age main sequence (TAMS). Although alternative initial conditions may lead to different binary histories, the resulting WD

[13] https://github.com/SihaoCheng/WD_models

evolutionary tracks generally occupy similar regions in the HR diagram.

During the early cooling phase, the proto-WD undergoes episodes of unstable hydrogen shell burning, which trigger recurrent hydrogen thermonuclear flashes (L. G. Althaus et al. 2013). These flashes appear in the red MESA WD track of Figure 9 as looping excursions before the star settles onto the quiescent final cooling branch. Green crosses indicate evolutionary ages spaced by 300 Myr, showing that the flash intervals are relatively brief compared with the much longer duration of the stable cooling phase. The WD models used for our parameter estimation (A. Bédard et al. 2020) include only the final, steady cooling sequences and omit the earlier flashing stages. This approach provides smoother grids for reliable interpolation of WD parameters; however, it also implies that if a WD is observed during a flash episode, its inferred mass can be significantly underestimated when fitted against these quiescent cooling tracks.

Given the relatively large uncertainties in the SED-derived WD parameters and the fact that several WD companions lie above the model cooling sequences or near the low-mass boundary, it is plausible that a small subset of systems are observed during or shortly after a flash stage. For these objects, the derived masses should be treated with caution. Nevertheless, the overall distribution strongly suggests that a substantial fraction of our WDMS candidates host genuinely low-mass, likely ELM WDs.

As shown in Figure 8, the MS companions are predominantly FGK-type stars. These companions are more luminous than M dwarfs, so the WD components must contribute sufficient UV flux to be detected in the combined photometry. This increases the likelihood of identifying WDs in the hydrogen-shell-flashing phase, since such WDs are intrinsically brighter than those already on the final cooling track at similar effective temperatures. In addition, the GALEX detection bias toward systems with stronger UV emission favors WDs with larger radii and hence lower masses, whereas more massive WDs, with smaller radii, require higher temperatures to produce a comparable UV excess.

This trend is illustrated in Figure 10, which compares the cumulative distributions of effective temperature for low-mass ($M \leqslant 0.4\,M_{\odot}$) and high-mass ($M \geqslant 0.8\,M_{\odot}$) WDs. The two groups show distinctly different temperature ranges: most low-mass WDs cluster between 10,000 and 20,000 K, with none exceeding 30,000 K, while high-mass WDs extend beyond 60,000 K. The median temperatures are 13,900 K and 19,600 K for the low- and high-mass subsets, respectively. This pattern supports the idea that our sample is biased toward either low-mass WDs with larger radii or massive WDs that are sufficiently hot to produce strong UV flux, consistent with the GALEX selection effects.

Recent work by N. Hallakoun et al. (2024) has revealed a notable deficit of massive WDs in Gaia astrometric binaries. Consistent trends have been reported in previous magnitude- or volume-limited WDMS samples, which show an enhanced fraction of low-mass WDs, although the degree of this excess varies among studies (e.g., A. Rebassa-Mansergas et al. 2021; P. K. Nayak et al. 2024; A. V. Sidharth et al. 2024; P. K. Nayak 2025; A. Rebassa-Mansergas et al. 2025). More recently, machine-learning classifications based on Gaia XP spectra (J. Li et al. 2025; X. Pérez-Couto et al. 2025) have provided independent support for this picture, and their WD components also show a strong concentration at low masses despite being paired primarily with late-type companions. Several factors may contribute to these trends. Gaia's angular resolution and scanning law naturally favor the detection of compact systems in which binary interaction has reduced the WD mass, producing a clear selection bias toward close orbital configurations and thereby increasing the prevalence of low-mass WDs in the observed population. In addition, WD progenitors with relatively massive MS companions may undergo enhanced mass loss during the RGB phase, producing systematically lower WD masses and further increasing the representation of low-mass WDs in the observed population.

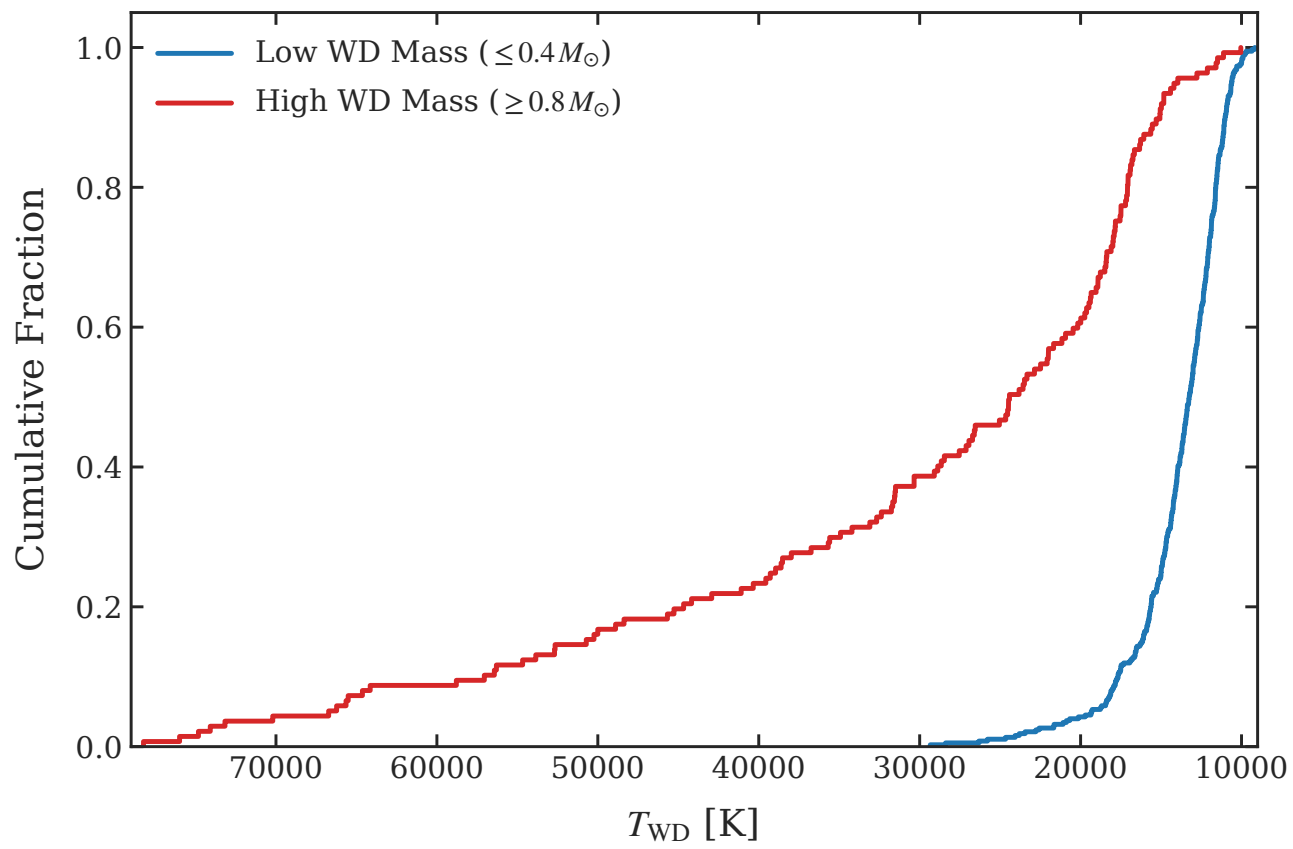

**Figure 10.** Cumulative distributions of effective temperatures for white dwarfs in the WDMS candidates, divided into two mass-selected subsets. Low-mass WDs ($M \leqslant 0.4\,M_{\odot}$) are shown in blue, while high-mass WDs ($M \geqslant 0.8\,M_{\odot}$) are shown in red. The divergence between the two curves illustrates the contrasting temperature distributions of the two groups.

Figure 11 presents the normalized distributions of key parameters for the WD components in our WDMS candidates, including effective temperature ($T_{\rm WD}$), radius ($R_{\rm WD}$), and mass ($M_{\rm WD}$). The figure also shows these distributions color coded by the spectral types of the corresponding MS companions, classified into FGK types as indicated in the top panel. The WD temperature distribution peaks near 15,000 K, with most objects cooler than 30,000 K. WD radii range from 0.006–0.17 $R_{\odot}$ (i.e., approximately 0.7–18 $R_{\oplus}$). According to B. Anguiano et al. (2022), companions with $R_{\rm WD} \gtrsim 25\,R_{\oplus}$ or $T_{\rm WD} \lesssim$ 9000 K are likely nondegenerate. Since all WDs in our candidates have $T_{\rm WD} >$ 9000 K and radii well within the degenerate range, these components can be regarded as bona fide WDs.

In the third panel of Figure 11, the radius distributions show a clear distinction depending on the spectral type of the MS companion. Among systems with F-, G-, and K-type stars, as the temperature of the MS companion decreases, the fraction of large-radius WD companions decreases significantly, with the fractions of WDs with $R_{\rm WD} > 0.04\,R_{\odot}$ being approximately 40%, 16%, and 4%, respectively. This trend indicates that WDMS binaries with earlier-type primaries are preferentially detected when the WD is more luminous at a given temperature, consistent with the UV-based selection bias discussed before.

The mass distribution in Figure 11 shows that systems with K-type MS companions exhibit a noticeable secondary peak near 0.5–0.6 $M_{\odot}$. This feature is more pronounced than in the F- and G-type subsamples and is consistent with expectations for wide, noninteracting WDMS binaries whose WD masses align with the field WD population. Across the full sample, the distribution remains dominated by low-mass WD companions, in agreement with the trends seen in Figure 9. The enhanced concentration of low-mass WDs likely reflects observational selection effects, including the UV sensitivity of GALEX and photometric detection limits, as well as systematic uncertainties inherent to SED-based mass estimates. Hydrogen shell flashes experienced by many low-mass He WDs before they settle onto their final cooling tracks may introduce additional biases in the inferred radii and luminosities.

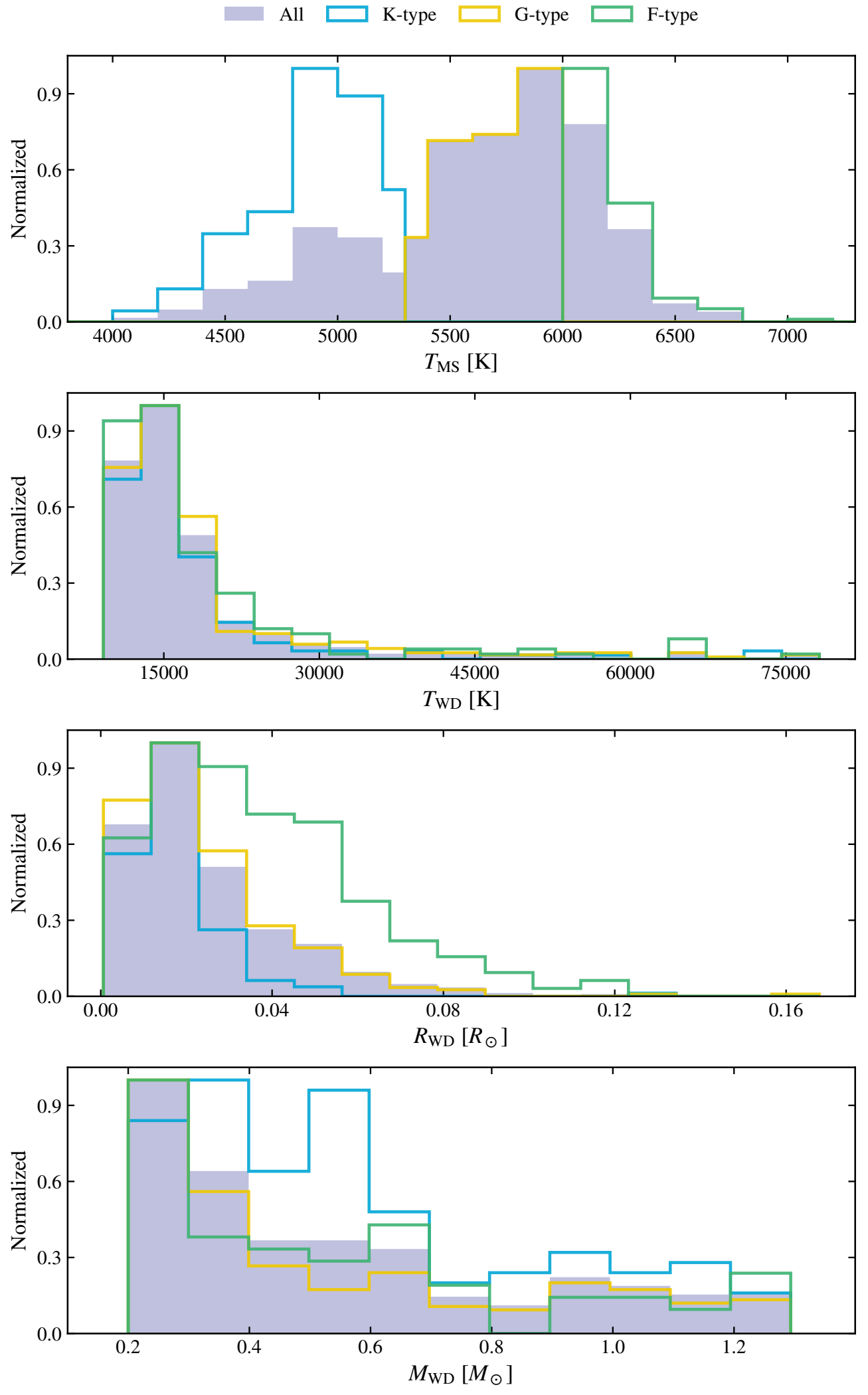

**Figure 11.** Normalized distributions of WD parameters for the WDMS candidate sample, separated by the spectral type of the MS companion as inferred from $T_{\rm MS}$. K-, G-, and F-type subsamples are shown in blue, yellow, and green, respectively, following the empirical scale of M. J. Pecaut & E. E. Mamajek (2013). The shaded purple histograms represent the full-sample distributions, and each subgroup is normalized independently. The top panel shows the $T_{\rm MS}$ distribution that defines the FGK classifications. The lower panels show the corresponding distributions of WD effective temperature ($T_{\rm WD}$), radius ($R_{\rm WD}$), and mass ($M_{\rm WD}$).

Future UV and spectroscopic observations will help refine the physical parameters of the WD companions. Time-domain monitoring of RV or photometric variability will also be essential for determining orbital periods and distinguishing close binaries from wide, noninteracting systems, thereby clarifying how binary evolution influences the observed WD mass distribution.

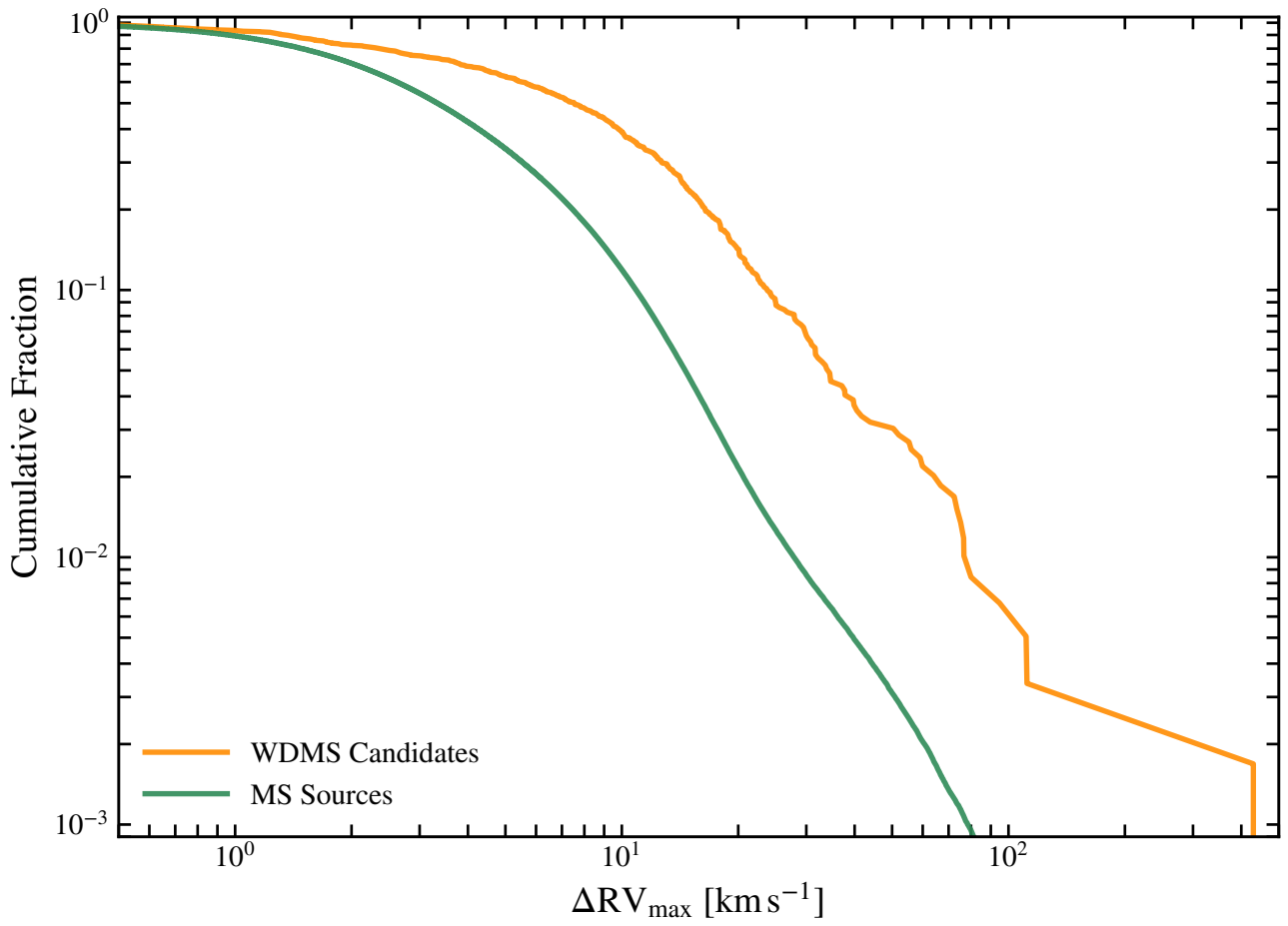

**Figure 12.** Cumulative distributions of the $\Delta{\rm RV}_{\rm max}$ for the WDMS candidates (orange curve) and the MS comparison sample (green curve) from Figure 1.

### *4.3. Radial-velocity Variations from LAMOST*

An advantage of the LAMOST survey is its multiepoch spectroscopic capability. Each released coadded spectrum is composed of several single exposures taken during the same observation to enhance the SNR and remove cosmic rays (Z.-R. Bai et al. 2017, 2021). To fully utilize this feature, we retrieved the corresponding single-epoch spectra for all sources in our sample and measured their RVs using the classical cross-correlation function method. As shown by Z.-R. Bai et al. (2021), the RV accuracy for F-, G-, and K-type stars is better than 10 km s$^{-1}$ when the SNR exceeds 15. Although most LAMOST targets have only one or two visits, the availability of multiepoch RVs still allows a meaningful statistical assessment of binarity and provides valuable insights into stellar and substellar multiplicity (G. Duchêne & A. Kraus 2013; Y. Chen et al. 2023; J. Liu et al. 2024).

For each WDMS candidate, we calculated the maximum RV variation, $\Delta{\rm RV}_{\rm max} = \max({\rm RV}) - \min({\rm RV})$, which serves as a sensitive tracer of potential orbital motion (C. Badenes & D. Maoz 2012). For comparison, the same calculation was performed for a reference sample of MS stars, corresponding to the sources remaining after Step 6 in Table 2 and shown as green points in Figure 1, using their multiepoch LAMOST spectra. To ensure reliable statistics, only targets with at least three spectroscopic epochs were included in both samples.

As shown in Figure 12, the cumulative distributions of $\Delta{\rm RV}_{\rm max}$ for the WDMS candidates (orange curve) and the MS comparison sample (green curve) differ markedly. The WDMS candidates are skewed toward larger $\Delta{\rm RV}_{\rm max}$ values, indicating shorter orbital periods. This behavior is consistent with expectations for systems containing low-mass WDs, which typically arise from either CE evolution or stable-RLOF in initially close binaries.

### *4.4. Comparison with Previous Literature*

Our WDMS candidate sample targets systems with FGK-type companions, occupying a distinct and comparatively underexplored region of parameter space. It therefore provides an important complement to the existing WDMS census. To evaluate how our candidates relate to previously identified populations, we compare them with five literature samples: a volume-limited WDMS sample (P. K. Nayak et al. 2024), a magnitude-limited WDMS catalog (A. Rebassa-Mansergas et al. 2025), two WDMS samples identified through artificial-intelligence methods using Gaia XP spectra (J. Li et al. 2025), and the compilation of WD binaries with FGK-type companions presented by J. A. Garbutt et al. (2024), which summarizes and analyzes previous related studies.

P. K. Nayak et al. (2024, hereafter N24) identified 93 WDMS systems within 100 pc using the Gaia CMD, and only one overlaps with our sample. A. Rebassa-Mansergas et al. (2025, hereafter RM25) published a larger catalog of 1312 WDMS binaries from Gaia DR3. None of these sources overlap with ours because the RM25 systems lie outside the MS region of the Gaia CMD, typically populating the gap between the WD cooling sequence and the MS. J. Li et al. (2025) applied a model-independent neural-network spectral model together with Gaussian-process classification to identify WDMS binaries from XP spectra. Their catalogs are dominated by M-type MS companions and therefore do not overlap with our FGK-selected sample. They provide two sets of WDMS candidates: a $\chi^2$-selected sample of 1649 systems (hereafter Li25$\chi^2$) and a larger quality-filtered sample of 30,131 systems identified through Gaussian-process classification (hereafter Li25GPC). J. A. Garbutt et al. (2024, hereafter G24) compiled WD+FGK binary candidates from three previous studies (S. G. Parsons et al. 2016; A. Rebassa-Mansergas et al. 2017; J. J. Ren et al. 2020), which did not provide detailed WD companion parameters. Using Gaia data, G24 derived stellar parameters for 206 WD+FGK systems, including systems with giant companions. Since G24 did not explicitly distinguish MS companions, we present the full G24 sample here rather than applying an additional WDMS-specific selection. Among these, only one overlaps with our sample.

In the top panel of Figure 13, the optical CMD distributions of these literature samples are compared with our WDMS candidates. The RM25 and Li25$\chi^2$ samples show strong spatial overlap and predominantly occupy the region between the WD cooling sequence and the MS. The Li25GPC sample lies mainly along the MS but extends toward redder colors, consistent with its late-type MS companions. The N24 systems appear more scattered, with only a few sources overlapping our WDMS locus. The G24 sample shows substantial overlap with our WDMS locus and also extends toward bluer colors, while many systems are distributed along the giant branch. In contrast, our WDMS candidates are more tightly concentrated along the FGK main-sequence region and are more numerous in this part of parameter space. Our candidates therefore populate the bluer portion of the FGK main-sequence region with a relatively large sample size, providing an important complementary set of WDMS binaries that has been comparatively underrepresented in previous studies.

The middle panel of Figure 13 shows the cumulative mass distributions of the WD components. Since the G24 sample includes both astrometric binaries and spectroscopic binaries, and the latter only provide lower limits on the WD companion masses through the binary mass function, our subsequent discussion of WD parameters includes only the astrometric binaries in G24, for which WD masses were estimated using the astrometric mass function. For the Li25$\chi^2$ and Li25GPC catalogs, WD masses are not provided directly. We therefore estimated them following the procedure described in J. Li et al. (2025), based on the parameters available in their tables. At the low-mass end, the cumulative rise of Li25$\chi^2$ closely matches that of our sample, consistent with the excess of low-mass WDs reported by J. Li et al. (2025). The Li25GPC sample shows the next steepest increase. Although these three samples occupy different regions in the optical CMD, they all exhibit a similar enhancement at low WD masses. This consistent behavior across CMD-distinct populations suggests that the prominence of low-mass WDs likely reflects an intrinsic property of WDMS binaries rather than arising purely from selection effects. In contrast, the G24 sample shows a more gradual increase at the lowest masses, but rises more rapidly in the 0.4–0.6 $M_\odot$ range and contains a relatively small fraction of high-mass WDs. The N24 sample contains a larger fraction of massive WDs ($M_{\rm WD} \gtrsim 0.8\,M_\odot$), whereas the RM25 sample shows the smallest fraction of low-mass WDs among the compared samples and does not exhibit a pronounced excess at the high-mass end; instead, its cumulative distribution rises more similarly to G24 in the intermediate-mass range.

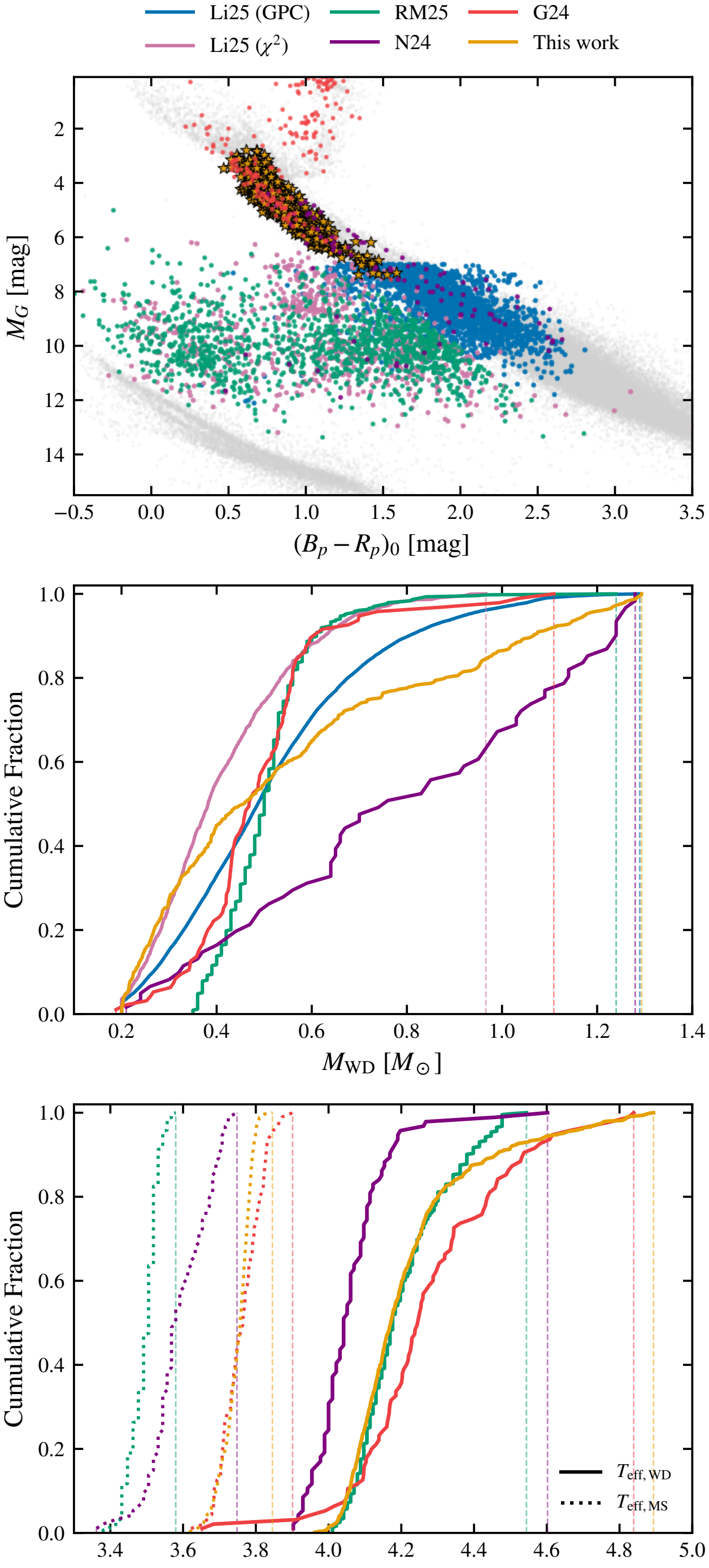

**Figure 13.** Comparison of our robust WDMS candidate sample with results from previous studies. Top panel: Gaia optical CMD showing our sample together with the WDMS samples of P. K. Nayak et al. (2024), A. Rebassa-Mansergas et al. (2025), J. Li et al. (2025; both the GPC- and $\chi^2$-selected samples), and the WD+FGK binary sample of J. A. Garbutt et al. (2024). Middle panel: cumulative distribution functions of WD companion masses for the compared samples. Bottom panel: cumulative distributions of effective temperatures for both the MS and WD components.

The bottom panel of Figure 13 compares the temperature distributions of both stellar components for the N24, RM25, and our sample; temperatures are not available for the Li25$\chi^2$ and Li25GPC systems. The MS companions in RM25 exhibit the lowest temperature distribution, while the G24 and our samples include systematically hotter MS companions. This offset is accompanied by a corresponding shift in the WD temperatures, and the upper limit of WD temperatures in RM25 remains lower than in our sample. This trend is expected, as detecting a WD next to a hotter MS star generally requires the WD itself to be intrinsically hotter in order to produce a detectable UV excess. For the G24 sample, the WD temperature distribution extends to higher temperatures, as expected, but also reaches down to values below ∼5000 K, with some WD temperatures even lower than those of their MS companions. Such cases may reflect differences in the temperature estimation methodology adopted in G24, and we do not pursue this issue further here. Around $T_{\rm WD} \sim$ 10,000–20,000 K, the cumulative temperature distributions of RM25 and our sample converge. However, because the MS stars in our sample are systematically hotter, the WDs in our systems must generally have larger radii and lower masses to produce a comparable UV excess. This naturally explains the higher fraction of low-mass WDs in our sample. This interpretation is consistent with the mass distributions shown in the middle panel and with Figure 10, where low-mass WDs preferentially occupy the lower-temperature regime. The MS temperature range in N24 is broader than in RM25, and the corresponding WD temperatures span a wider interval as well.

Although S. G. Parsons et al. (2016), A. Rebassa-Mansergas et al. (2017), and J. J. Ren et al. (2020), as summarized by G24, presented samples of WD+AFGK binaries, these works do not report WD temperatures, masses, or other derived parameters and are therefore not included in the above comparative analysis. While a nonnegligible subset of their systems overlaps with our sample, the majority of our WDMS candidates represent newly identified binaries. By providing homogeneous physical parameters for both stellar components, our study not only enlarges the current census but also establishes a systematically characterized population of WDMS binaries within this region of parameter space.

## 5. Summary and Conclusions

In this study, we identified a final sample of 654 robust WDMS candidates with FGK-type companions, selected from an initial set of 772 UV-excess sources, among which 759 yielded reliable binary SED fits. The selection combines high-precision astrometry and photometry from Gaia DR3 with UV measurements from GALEX GR6+7. By examining the locations of sources in the UV and optical CMDs, the relation between UV color and LAMOST spectroscopic temperature, and their positions in these diagnostic planes, we effectively separated unresolved WDMS binaries from single MS stars and other contaminants.

We performed binary SED fitting for all WDMS candidates, deriving posterior estimates of the effective temperatures and radii of both components, along with a common distance and extinction. Because the MS companion dominates the optical-to-infrared SED, the WD parameters should be interpreted with caution. Nevertheless, the MS temperatures agree closely with LAMOST spectroscopic estimates. The MS companions consist of approximately 24% F-type, 52% G-type, and 24% K-type stars, and their temperatures cluster around $T_{\rm MS} \sim 6000$ K. The WD components are typically hotter, with $T_{\rm WD} \sim 15{,}000$ K.

After assessing possible contamination from chromospherically active stars, chance alignments, and hot subdwarfs, and after verifying that the companions are consistent with the main-sequence evolutionary phase based on isochrone classification, we obtained a refined catalog of 654 reliable WDMS binaries. The MS companions are dominated by G-type stars. Moreover, as the MS companion temperature increases from K- to F-type, the associated WDs tend to exhibit larger radii, consistent with the UV-selection bias favoring more luminous WD components.

Using the evolutionary cooling models of A. Bédard et al. (2020), we derived WD masses from the SED-inferred temperatures and luminosities. The resulting WD mass distribution is strongly skewed toward low masses: most systems contain WDs in the 0.2–0.4 $M_\odot$ range, including a substantial number of ELM WDs ($\lesssim 0.3\,M_\odot$), consistent with expectations for products of binary mass loss. The WD population in our sample is therefore characterized by relatively large radii and low masses, while the more massive WDs correspondingly exhibit higher effective temperatures. The cumulative distribution of LAMOST $\Delta{\rm RV}_{\rm max}$ values is shifted toward larger amplitudes relative to a comparison sample of MS stars, reinforcing the interpretation that a significant fraction of our systems is close binaries. These results collectively support a scenario in which many of the WDMS binaries identified here have undergone substantial binary interaction, leading to the formation of low-mass WDs.

The observed WD parameter distributions reflect the combined effects of selection biases—favoring hotter, more luminous, and thus larger-radius WDs—and possible mass underestimation for systems observed during transient hydrogen-shell-flash phases before reaching the final cooling branch. The WD properties in our analysis rely primarily on GALEX UV photometry, and most systems lack multiepoch or high-resolution spectroscopy, so their orbital parameters remain largely unconstrained. It also remains possible that the intrinsic evolutionary pathways of WDMS binaries play a fundamental role in producing the ELM WDs revealed in our sample.

Future high-resolution spectroscopic observations, together with time-domain monitoring, will be essential for confirming binarity, refining the physical parameters of the WD components, and clarifying the evolutionary histories of FGK+WD binaries. Such observations will further enable a more comprehensive understanding of the population demographics of WDMS systems and their role in binary evolution.

**Acknowledgments**

This work is supported by the National Natural Science Foundation of China (grant Nos. 12125303, 12288102, 12273056, 12090041, 12090040, and 12090043), the National Key R&D Program of China (grant Nos. 2021YFA1600401, 2021YFA1600403, and 2022YFA1603002), the International Centre of Supernovae, Yunnan Key Laboratory (grant No. 202505AV340004), the Yunnan Revitalization Talent Support Program-Science & Technology Champion Project (grant No. 202305AB350003), the New Cornerstone Science Foundation through the XPLORER PRIZE, and the Strategic Priority Research Program of the Chinese Academy of Sciences (grant Nos. XDB1160000 and XDB1160200).

H.Y. and Z.B. are supported by the National Key R&D Program of China (grant No. 2023YFA1607901). H.Y. also acknowledges support from the Youth Innovation Promotion Association of the CAS (grant No. 20200060), the National Natural Science Foundation of China (grant No. 11873066), and the Strategic Priority Program of the Chinese Academy of Sciences (grant No. XDB1160302).

J.X. is supported by the National Natural Science Foundation of China (grant No. 12303106) and the China Postdoctoral Science Foundation (CPSF; grant No. GZC20232976). J.L. is supported by the Yunnan Fundamental Research Projects (grant No. 202501CF070016) and the Young Talent Project of the Yunnan Revitalization Talent Support Program. M.Z. is supported by the National Natural Science for Youth Foundation of China (grant No. 12503091).

Guoshoujing Telescope (the Large Sky Area Multi-Object Fiber Spectroscopic Telescope LAMOST) is a National Major Scientific Project built by the Chinese Academy of Sciences. Funding for the project has been provided by the National Development and Reform Commission. LAMOST is operated and managed by the National Astronomical Observatories, Chinese Academy of Sciences.

This work presents results from the European Space Agency (ESA) space mission Gaia. Gaia data are being processed by the Gaia Data Processing and Analysis Consortium (DPAC). Funding for the DPAC is provided by national institutions, in particular the institutions participating in the Gaia MultiLateral Agreement (MLA). The Gaia mission website is https://www.cosmos.esa.int/gaia. The Gaia archive website is https://archives.esac.esa.int/gaia. We acknowledge use of the VizieR catalog access tool, operated at CDS, Strasbourg, France, and of Astropy, a community-developed core Python package for Astronomy (Astropy Collaboration et al. 2013, 2018, 2022).

*Software*: ARIADNE (J. I. Vines & J. S. Jenkins 2022), astropy (Astropy Collaboration et al. 2013, 2018, 2022), dustmaps (G. Green 2018), matplotlib (J. D. Hunter 2007), MESA version r23.05.1 (B. Paxton et al. 2011, 2013, 2015, 2018, 2019; A. S. Jermyn et al. 2023), numpy (C. R. Harris et al. 2020), pandas (W. McKinney 2010), PyMultiNest (J. Buchner et al. 2014), scipy (P. Virtanen et al. 2020), VizieR (F. Ochsenbein et al. 2000).

## Appendix
## Catalogs of WDMS Candidates

In this appendix, we provide supplementary material related to the construction of our WDMS candidate catalog. Table A1 lists the 772 UV-excess sources identified in Section 2.3, together with the prior values adopted for the subsequent SED fitting. Table A2 presents the 759 WDMS binary candidates with well-fitted SEDs, including atmospheric parameters from LAMOST spectroscopy, the fitted binary parameters from the SED modeling, and the estimated stellar masses and ages. Quality flags are provided to enable users to select the 654 highly reliable WDMS candidates defined in this work. Only a subset of entries is shown here for illustration; the complete machine-readable catalogs are available at doi:10.5281/zenodo.17799534.

**Table A1**
WDMS Candidates Exhibiting Ultraviolet Excess Located on the Optical Main Sequence

| R.A. (deg) | Decl. (deg) | LAMOST UID | Gaia ID | GALEX ID | $T_{\mathrm{eff}}$ (K) | $\log g$ (dex) | [Fe/H] (dex) | Distance (pc) | $\mathrm{EBV}_{\mathrm{Bayestar}}$ (mag) | $\mathrm{EBV}_{\mathrm{SFD}}$ (mag) | flag_extinction |
|---|---|---|---|---|---|---|---|---|---|---|---|
| 0.44755 | 27.45924 | G17199448424129 | 2853975398479650688 | 6376756377363154532 | 6069.8 ± 18.4 | 4.31 ± 0.03 | −0.28 ± 0.02 | 1238.29 ± 57.43 | 0.070 ± 0.004 | 0.048 | Bayestar |
| 0.62738 | 15.93837 | G17083494267623 | 2772219901429946240 | 6376052689921378750 | 6075.5 ± 30.8 | 4.17 ± 0.04 | 0.18 ± 0.03 | 725.50 ± 11.77 | 0.060 ± 0.010 | 0.039 | Bayestar |
| 0.68161 | 27.03744 | G17199412152837 | 2853902521474600064 | 6376756452525083713 | 5513.2 ± 49.3 | 4.60 ± 0.07 | 0.17 ± 0.05 | 644.56 ± 9.56 | 0.070 ± 0.000 | 0.047 | SFD |
| 1.54897 | 19.21189 | L17080846932058 | 2797983348654791168 | 6375982371643068298 | 6127.2 ± 67.7 | 4.61 ± 0.11 | −0.08 ± 0.07 | 696.45 ± 12.08 | 0.060 ± 0.008 | 0.055 | Bayestar |
| 1.71704 | 1.49735 | G17046519287737 | 2738763484879678848 | 6381013755105382445 | 5794.0 ± 26.1 | 4.40 ± 0.04 | 0.35 ± 0.02 | 207.39 ± 1.29 | 0.000 ± 0.011 | 0.030 | SFD |
| 1.82985 | 26.28285 | G17199819250237 | 2850625976823955328 | 6375982306144818141 | 5258.4 ± 50.8 | 4.41 ± 0.05 | −0.05 ± 0.02 | 137.90 ± 0.45 | 0.000 ± 0.000 | 0.036 | SFD |
| 10.15195 | 16.16268 | L17273485485844 | 2782138909357390336 | 6380943346632755644 | 5228.9 ± 51.4 | 4.70 ± 0.07 | −0.09 ± 0.04 | 491.94 ± 5.34 | 0.020 ± 0.004 | 0.041 | Bayestar |
| 10.45433 | 27.91172 | G17234699082764 | 2809588973059319680 | 6376721250973123550 | 5058.3 ± 35.8 | 4.71 ± 0.05 | −0.02 ± 0.03 | 385.98 ± 3.66 | 0.060 ± 0.010 | 0.045 | Bayestar |
| 10.57235 | 21.92057 | G17278635546239 | 2802538835782232064 | 6380943323010435622 | 5496.2 ± 49.5 | 4.28 ± 0.07 | −0.46 ± 0.05 | 349.45 ± 2.28 | 0.070 ± 0.014 | 0.045 | Bayestar |
| 100.24828 | 36.05456 | G16092624296001 | 942084862878302592 | 6373800953458460183 | 5850.6 ± 18.2 | 4.41 ± 0.02 | 0.25 ± 0.01 | 192.06 ± 0.65 | 0.003 ± 0.003 | 0.131 | Bayestar |

**Note.** The table lists the R.A. and decl. (J2000 coordinates), LAMOST spectrum unique identifiers (UIDs), and corresponding Gaia and GALEX source IDs. For each object, we provide the atmospheric parameters and uncertainties from LAMOST spectroscopy. Distance is derived from the Gaia parallax. The extinction EBV is given from two dust maps (G. M. Green et al. 2019) and D. J. Schlegel et al. (1998), and flag_extinction indicates which extinction we used as prior in the SED fitting.

(This table is available in its entirety in machine-readable form in the online article.)

**Table A2**
WDMS Candidates with Good Binary SED Fits

| R.A. (deg) | Decl. (deg) | LAMOST UID | Gaia ID | GALEX ID | $T_{\rm eff}$ (K) | $\log g$ (dex) | [Fe/H] (dex) | $R^{+}_{\rm HK}$ | $T_{\rm MS}$ (K) | $R_{\rm MS}$ ($R_\odot$) | $T_{\rm WD}$ (K) | $R_{\rm WD}$ ($R_\odot$) | Distance (pc) | $A_V$ (mag) | $Vgf_{\rm b}$ | $M_{\rm MS}$ ($M_\odot$) | $\rm Age_{MS}$ (Gyr) | EEP | $M_{\rm WD}$ ($M_\odot$) | $\rm Age_{WD}$ (Gyr) | flag_Xray | flag_chance | flag_noms |
|---|---|---|---|---|---|---|---|---|---|---|---|---|---|---|---|---|---|---|---|---|---|---|---|
| 0.4476 | +27.459 | G17199448424129 | 2853975398479650688 | 6376756377363154532 | 6070 ± 18 | 4.31 ± 0.03 | −0.28 ± 0.02 | 0.11 | $6287^{+43}_{-52}$ | $1.28^{+0.04}_{-0.02}$ | $13{,}370^{+1915}_{-643}$ | $0.030^{+0.009}_{-0.007}$ | $1192^{+24}_{-17}$ | $0.166^{+0.013}_{-0.005}$ | 2.58 | 1.08 | 0.10 | 243.6 | | | False | False | False |
| 0.6274 | +15.938 | G17083494267623 | 2772219901429946240 | 6376052689921378750 | 6076 ± 31 | 4.17 ± 0.04 | 0.18 ± 0.03 | 0.18 | $5763^{+172}_{-49}$ | $1.68^{+0.06}_{-0.08}$ | $14{,}887^{+1926}_{-1289}$ | $0.043^{+0.012}_{-0.011}$ | $724^{+10}_{-5}$ | $0.141^{+0.022}_{-0.007}$ | 5.39 | 1.21 | 5.13 | 411.6 | | | False | False | False |
| 1.5490 | +19.212 | L17080846932058 | 2797983348654791168 | 6375982371643068298 | 6127 ± 68 | 4.61 ± 0.11 | −0.08 ± 0.07 | -0.01 | $5931^{+18}_{-16}$ | $1.26^{+0.01}_{-0.01}$ | $18{,}458^{+1738}_{-514}$ | $0.014^{+0.002}_{-0.001}$ | $693^{+3}_{-2}$ | 0.155 ± 0.004 | 0.61 | 1.09 | 0.02 | 191.9 | 0.56 | 0.07 | False | False | True |
| 1.7170 | +1.497 | G17046519287737 | 2738763484879678848 | 6381013755105382445 | 5794 ± 26 | 4.40 ± 0.04 | 0.35 ± 0.02 | 0.15 | $5630^{+34}_{-55}$ | $1.10^{+0.02}_{-0.01}$ | $65{,}513^{+4551}_{-5272}$ | $0.001^{+0.000}_{-0.000}$ | $208^{+1}_{-1}$ | $0.081^{+0.023}_{-0.024}$ | 0.88 | 1.06 | 5.13 | 372.9 | | | False | False | False |
| 1.8299 | +26.283 | G17199819250237 | 2850625976823955328 | 6375982306144818141 | 5258 ± 51 | 4.41 ± 0.05 | −0.05 ± 0.02 | 0.43 | $4823^{+53}_{-16}$ | $1.08^{+0.01}_{-0.03}$ | $11{,}251^{+199}_{-239}$ | $0.012^{+0.003}_{-0.001}$ | $138^{+0}_{-0}$ | $0.056^{+0.005}_{-0.009}$ | 6.39 | 0.76 | 5.67 | 331.1 | 0.66 | 0.53 | False | False | False |

**Note.** Columns that are identical to those in Table A1 (e.g., coordinates, LAMOST UIDs, Gaia/GALEX source IDs, atmospheric parameters, and distances) carry the same definitions as described there. In addition, this table includes the chromospheric activity index $R^{+}_{\rm HK}$ measured from the Ca II H and K lines, the binary SED fitting results with the visual goodness-of-fit parameter $Vgf_{\rm b}$, and the derived stellar masses, ages, and EEP values for the MS companions interpolated from MIST isochrones. WD masses and cooling ages are estimated from white-dwarf evolutionary cooling models. The flags `FLAG_Xray`, `FLAG_CHANCE`, and `FLAG_NOMS` mark sources considered potential contaminants and excluded accordingly.

(This table is available in its entirety in machine-readable form in the online article.)

## ORCID iDs

Mingkuan Yang (杨明宽) https://orcid.org/0009-0000-6595-2537
Hailong Yuan (袁海龙) https://orcid.org/0000-0002-4554-5579
Xiaozhen Yang (杨肖振) https://orcid.org/0009-0006-7506-1299
Zhongrui Bai (白仲瑞) https://orcid.org/0000-0003-3884-5693
Yuji He (何玉吉) https://orcid.org/0009-0008-0146-118X
Jianping Xiong (熊建萍) https://orcid.org/0000-0003-4829-6245
Jiao Li (李蛟) https://orcid.org/0000-0002-2577-1990
Mengxin Wang (汪梦欣) https://orcid.org/0009-0009-6931-2276
Yiqiao Dong (董义乔) https://orcid.org/0000-0002-6312-4444
Ziyue Jiang (蒋子悦) https://orcid.org/0009-0009-9065-1846
Qian Liu (刘倩) https://orcid.org/0009-0009-0971-5081
Ganyu Li (李甘雨) https://orcid.org/0009-0009-0171-1553
Ming Zhou (周明) https://orcid.org/0000-0003-1487-0093
Haotong Zhang (张昊彤) https://orcid.org/0000-0002-6617-5300
Xuefei Chen (陈雪飞) https://orcid.org/0000-0001-5284-8001